\documentclass[fleqn,usenatbib]{mnras}

\usepackage{newtxtext,newtxmath}

\usepackage[T1]{fontenc}
\usepackage[utf8]{inputenc}

\DeclareRobustCommand{\VAN}[3]{#2}
\let\VANthebibliography\thebibliography
\def\thebibliography{\DeclareRobustCommand{\VAN}[3]{##3}\VANthebibliography}

\usepackage{graphicx}	
\usepackage{amsmath}	
\usepackage{physics}    
\usepackage{ulem}       
\usepackage{bigints}
\usepackage{fix-cm}

\usepackage{multirow}
\usepackage{booktabs}

\newcommand{\vect}[1]{\boldsymbol{#1}}
\renewcommand{\vec}[1]{\boldsymbol{#1}}

\newcommand{\equ}[1]{eq.~(\ref{eq:#1})}
\newcommand{\equs}[1]{eqs.~(\ref{eq:#1})}
\newcommand{\equm}[1]{(\ref{eq:#1})}

\newcommand{\se}[1]{Section~\ref{sec:#1}}
\newcommand{\ses}[1]{Sections~\ref{sec:#1}}

\newcommand{\fig}[1]{Fig.~\ref{fig:#1}}
\newcommand{\figs}[1]{Figs.~\ref{fig:#1}}
\newcommand{\figss}[1]{\ref{fig:#1}}

\newcommand{\tab}[1]{Table~\ref{tab:#1}}
\newcommand{\be}{\begin{equation}}
\newcommand{\ee}{\end{equation}}
\newcommand{\bea}{\begin{eqnarray}}
\newcommand{\eea}{\end{eqnarray}}

\newcommand{\no}{\noindent}

\newcommand{\Msun}{M_\odot}

\newcommand{\ifm}[1]{\relax\ifmmode#1\else$\mathsurround=0pt #1$\fi}
\newcommand{\kms}{\ifmmode\,{\rm km}\,{\rm s}^{-1}\else km$\,$s$^{-1}$\fi}

\newcommand{\Mpc}{\,{\rm Mpc}}
\newcommand{\kpc}{\,{\rm kpc}}

\newcommand{\Gyr}{\,{\rm Gyr}}

\newcommand{\Myr}{\,{\rm Myr}}

\newcommand{\ltsima}{$\; \buildrel < \over \sim \;$}
\newcommand{\lsim}{\lower.5ex\hbox{\ltsima}}
\newcommand{\gtsima}{$\; \buildrel > \over \sim \;$}
\newcommand{\gsim}{\lower.5ex\hbox{\gtsima}}

\def\la{\langle}

\def\cmc{\,{\rm cm}^{-3}}

\def\nH{n_{\rm H}}

\def\M*{M_{\rm *}}

\def\Kdegree{{\rm K}}

\def\Pi{\varpi_{_{\rm I}}}
\def\Gauss{\rm G}

\usepackage{pifont}

\usepackage{scalerel}

\title[B-field evolution]{Magnetic field evolution in feedback-dominated galaxies}

\author[Y. S. Lu et al.]{\parbox[t]{\textwidth}{Yue Samuel Lu,$^{1,2}$\thanks{E-mail: yul232@ucsd.edu}
Dušan Kereš,$^{1, 2}$
Philip F. Hopkins, $^{3}$ and
Sam B. Ponnada$^{4}$
}
\vspace{0.2cm}\\%
$^{1}$Department of Physics, University of California, San Diego, La Jolla, CA 92093, USA\\%
$^{2}$Department of Astronomy and Astrophysics, University of California, San Diego, La Jolla, CA 92093, USA\\%
$^{3}$TAPIR, California Institute of Technology, Mailcode 350-17, Pasadena, CA 91125, USA \\%
$^{4}$Department of Physics and Astronomy, Chalmers University of Technology, Gothenburg, Sweden \\%
}
\date{Accepted XXX. Received YYY; in original form ZZZ}

\pubyear{2026}

\begin{document}
\label{firstpage}
\pagerange{\pageref{firstpage}--\pageref{lastpage}}
\maketitle

\begin{abstract}
Over the last decade, magnetic fields (B-fields) have become a common ingredient in galaxy formation simulations. Although seemingly unimportant in matching observational constraints such as the $M_*-M_{\rm halo}$ relations, the B-field influences the circumgalactic medium (CGM) gas and determines cosmic ray (CR) propagation, making it an indispensable non-thermal component. Nevertheless, the intricate physics of B-fields, particularly how they evolve to the observed $\mu$G-level field, remains largely elusive. In this work, we systematically address the evolution of B-fields over cosmic time using the FIRE-2 (Feedback In Realistic Environments) magnetohydrodynamics (MHD) simulations. We present results for both Milky Way (MW)-mass and dwarf galaxies, with and without CRs. We find that CRs have little impact on B-field properties, while the halo mass controls the saturated B-field strength at $z\sim 0$. However, all runs reach a halo-independent magnetic-to-turbulent ratio of $f_{\rm B}\sim 1-10\%$, though the methods used to measure turbulence could affect this value. We identify a three-stage evolutionary course for the B-field in tandem with galaxy formation: i. expansion, ii. amplification, and iii. saturation. The feature is universal, though the timings depend on the initial location and star formation history of the gas. Flux-freezing is the predominant driver for the first and last stages, exhibiting power relations $B\sim \rho^{2/3}$ and $\rho^{1/2}$, although the latter could also be explained by dynamical and energetic equilibria. The intermediate amplification stage, potentially driven by ``dynamos’’, is fuelled by structure formation, feedback, and gas turbulence. We conjecture a numerically triggered, turbulence-driven amplification process to explain this phenomenon.
\end{abstract}




\section{Introduction}
\label{sec:intro}
Magnetic fields (B-fields) permeate the Universe ubiquitously across all scales: from planets to stars and accretion disks around supermassive black holes (SMBHs), from the interplanetary to the interstellar medium (ISM), and  from the circumgalactic medium (CGM) to the intergalactic medium (IGM). The current observed magnetic field strength is of $\mu \Gauss $-level in the ISM of the Milky-Way (MW) \citep{parker1971generation}. The B-fields in galaxies are of particular interest, as they serve as the origin of B-fields in stars \citep{mestel1960star}, determine cosmic ray (CR) propagation \citep{jokipii1966cosmic}, and might affect some stages of star formation \citep{federrath2016magnetic, kaalva2026evolution}.

Both the physical origin and the course of evolution of galactic B-field are mostly unclear. Given the uncertain origins, it is a common practice in modern cosmological simulations to set the initial B-fields, or the ``seed field'', to be some weak ($\sim {\rm n}\Gauss$) imposed fields (e.g. a homogeneous field). However, many works have shown that the choice of seed fields and the seeding methods have little to no impact on the properties of galactic B-fields at low-$z$ \citep{martin2021unravelling, ramesh2026magnetogenesis}, being able to produce observables consistent with observations (e.g., see \citet{pakmor2013simulations, ponnada2022magnetic}). Some theories have proposed possibilities of how B-fields were seeded, including the Biermann battery \citep{attia2021cosmological}, the dust battery \citep{soliman2025dust}, and supernova or black hole injection \citep{martin2021unravelling, ramesh2026magnetogenesis}. Or, the B-field could have been completely ``primordial'', arising in the early Universe without any astrophysical processes \citep{ratra1992cosmological, barrow2011cosmological}, though the distinction between primordial and astrophysically originated B-fields is somewhat nebulous \citep{garaldi2021magnetogenesis}, as we have minimal observational evidence for B-fields at such high redshifts.

Galactic B-fields are highly turbulent and exhibit complex spatial structures, as revealed by both observations and simulations \citep{moss2012multiscale, beck2016magnetic, van2021effect, su2018stellar}. Only a very limited number of observational methods are available to observe the galactic B-fields, mostly indirect, including Faraday rotation measure (RM, \citet{prochaska2020constraining}), dust polarization \citep{hoang2008radiative, hoang2014grain}, Zeeman effects \citep{basu2013magnetic}, and radio synchrotron emission \citep{basu2013magnetic, beck2015magnetic}. Synthetic observations have been generated from simulations with B-field implementations to compare with observations, e.g., \citet{pakmor2018faraday} for RM, and \citet{pfrommer2022simulating, ponnada2023synchrotron} for radio synchrotron emission.

In general, galactic B-fields are dynamically and energetically unimportant. Although observed to be roughly in equipartition with the thermal, turbulent, and CR energies in the solar neighbourhood \citep{boulares1990galactic, beck2016magnetic}, galaxy simulations showed that the strength of B-fields could drop drastically in the CGM, making up $\lsim 10\%$ of the turbulent and thermal energy in MW-mass galaxies, while the values of the magnetic-to-turbulent ratio can range from $\lsim 1\%$ to $\sim 20\%$ depending on the code and feedback recipes \citep{pakmor2020magnetizing, rieder2016small-I, ponnada2022magnetic, marinacci2018first}. However, B-fields may be energetically important in dwarf galaxies, where less gas is retained and weaker turbulence is present due to their low masses \citep{kepley2011magnetic, martin2025pandora}.  

Although B-fields have already been included in many cosmological galaxy simulations with various feedback models, both large-scale and zoom-in, how the galactic B-field was amplified during the course of galaxy formation and evolution is still a total mystery. Turbulence of gas, feedback processes, and large-scale inflow/outflow could all potentially be responsible for the amplification of the B-field in galaxy evolution. During the early stage of galaxy evolution, \underline{dynamos}\footnote{Although the term ``dynamo'' is widely used with clear definitions in some fields (e.g. fusion plasma), in the context of galaxies, its semantic and physical meanings are somewhat nebulous, often referring to little more than ``something responsible for B-field amplification''. This ambiguity largely stems from the fact that the actual physical mechanisms driving B-field growth remain uncertain, which is precisely what we explore in this work.} are thought to be the primary amplification source, and later stage amplification is mostly dominated by flux-freezing. This broad trend was seen in multiple simulation works, including the Auriga simulation project (run with the \texttt{AREPO} code, \citet{pakmor2020magnetizing, pakmor2024magnetic}), simulations run with \texttt{RAMSES} code \citep{rieder2016small-I, rieder2017small-II, rieder2017small-III, martin2018three}, simulations run with \texttt{ENZO} code \citep{butsky2017ab}, and, as will be shown in this work, FIRE simulations (run with \texttt{GIZMO} code). On the other hand, numerical effects usually have non-negligible impacts on the B-field growth (especially in particle codes, see the recent work \citet{tomida2026systematic}), as the divergence cleaning schemes could create complications and affect the actual B-field physics \citep{hopkins2016accurate, hopkins2016constrained}. Among the aforementioned works, some used idealized and some used cosmological simulations, and the feedback models included were also different. Nevertheless, they all more or less reached the similar conclusion that feedback is indeed critical to B-field amplification. 

Among many simulations with feedback models, perhaps one of the most timely implementation is the Feedback In Realistic Environment (FIRE) project\footnote{FIRE project website: \hyperlink{http://fire.northwestern.edu}{http://fire.northwestern.edu}} \citep{hopkins2014galaxies, hopkins2018fire, hopkins2023fire}. We summarize the simulation methods, including the implementation of B-fields, in \se{methods}. \citet{su2017feedback} implemented the B-field in tandem with FIRE-2 feedback physics, and \citet{su2018stellar} was the first FIRE-related work to discuss the structure of the B-field in the galaxy. Further works studying B-field structure and corresponding observables include \citet{ponnada2022magnetic, ponnada2023synchrotron}. Although \citet{ponnada2022magnetic} performed a thorough analysis on the structure of the B-field in $z=0$, ${\rm L}_*$-- galaxies and generated multiple synthetic observations to compare with real observations, they did not explore the evolution of B-field. This provides the motivation for this work.

In this paper, we aim to draw a more comprehensive picture of galactic B-fields \footnote{Since this paper will focus exclusively on B-fields in galaxies, we will not make this distinction hereafter and all mentions of  ``B-fields'' will mean galactic B-fields, unless stated explicitly.} in feedback-dominated galaxies over the course of galaxy evolution. We will focus on the general amplification process and aim to reveal potential connection to dynamos. 

This paper is organized as follows. In \se{methods} we introduce the FIRE-2 simulations, with a particular focus on the implementation of B-fields and CRs, and present the simulations we select for this work. In \se{properties} we present the properties of the B-field in our galaxies at low-$z$. In \se{amplification} we use Eulerian (fixed-volume) and Lagrangian (particle tracing) methods to address the amplification of the B-field over cosmic time in tandem with galaxy evolution. In \se{discussion} we present some robustness tests and conjecture a potential dynamo origin, and in \se{conclusions} we conclude.

\section{Methods}
\label{sec:methods}
In this section we briefly discuss the methodological setup of this exercise. In \se{fire-2} we review the core physics implemented in the FIRE-2 feedback models. In \se{B-fields and CRs} we recapitulate the main physics of B-fields and CRs and how they were evolved in FIRE-2 simulations. In \se{selected sims} we introduce the specific FIRE-2 simulations we analyzed in this paper.

\subsection{FIRE-2 simulations}\label{sec:fire-2}
We use simulations evolved with the FIRE-2 method for star formation and feedback \citep{hopkins2018fire}, which is an updated version of the FIRE-1 model \citep{hopkins2014galaxies}. Here, we briefly describe the FIRE-2 methods and refer the readers to the original references for further details. The simulations are run with the \texttt{GIZMO} code \citep{hopkins2015new}, with a meshless finite-mass MFM solver (a mesh-free finite-volume Lagrangian Godunov method). All simulations presented in this work include ideal magnetohydrodynamics (MHD), thermal conduction and viscosity \citep{hopkins2020but}, and gravity with adaptive Lagrangian force softening for gas. These are cosmological ``zoom-in'' simulations, evolving a large box from $z \gsim 100$ with resolution concentrated on a $\sim 1 - 10 \Mpc$ comoving volume around a target halo of interest.

Radiative cooling, star formation, and stellar feedback mechanisms are all included following the FIRE-2 implementation \citep{hopkins2018fire}. Gas cooling is followed from $10-10^{10} \Kdegree$ including Compton, free-free, metal line, fine-structure, atomic, and molecular processes. We also follow photoelectric and photoionization gas heating by both local sources
and an uniform metagalactic background including the effect of self-shielding \citep{faucher2009new}. As for stars, star formation is allowed when all the following criteria are satisfied: the gas is locally self-gravitating \citep{hopkins2013meaning, grudic2018feedback}, self-shielding, Jeans-unstable, molecular, and dense ($n >1000 \cmc$). Stars then evolve according to standard star evolution tracks, assuming Kroupa IMF \citep{kroupa2011stellar} and explicitly accounting for the mass, metals, momentum, and energy injected via individual supernovae (SNe, type Ia \& II), O/B and AGB-star mass-loss \citep{hopkins2018model}, and their radiation \citep{hopkins2020radiative}.

\begin{table*}
    \resizebox{\textwidth}{!}
        {\begin{tabular}{c c c c c c c}
        \toprule
        \multicolumn{1}{c}{Target halo} & \multicolumn{1}{c}{Basic physics} & \multicolumn{1}{c}{$m_{\rm gas \, cell}$} & \multicolumn{1}{c}{$\mathrm{R}_{\mathrm{vir}}$} & \multicolumn{1}{c}{$\mathrm{M}_{\mathrm{vir}}$} & \multicolumn{1}{c}{$\mathrm{M}_{\star}(<0.1 R_{\rm vir})$} & \multicolumn{1}{c}{Notes} \\
        \multicolumn{1}{c}{} & \multicolumn{1}{c}{} & \multicolumn{1}{c}{$[\Msun]$} & \multicolumn{1}{c}{[pkpc]} & \multicolumn{1}{c}{$[\mathrm{M}_{\odot}]$} & \multicolumn{1}{c}{$[\mathrm{M}_{\odot}]$} & \multicolumn{1}{c}{} \\
        \midrule

        \multicolumn{7}{c}{\textbf{MW-mass galaxies}} \\
        \addlinespace

        \multirow{2}{*}{\texttt{m12i}}     & FIRE-2 MHD+    & \multirow{4}{*}{$\approx 7\times 10^3$}    & \multirow{2}{*}{275}  & \multirow{2}{*}{$1.0\times 10^{12}$} & $7\times 10^{10}$  & \multirow{2}{*}{``Latte'' halo, large, late-forming disk}         \\
             & FIRE-2 CR-CD   &      &    &    & $3\times 10^{10}$ &    \\
        \addlinespace

        \multirow{2}{*}{\texttt{m12f}}    & FIRE-2 MHD+    &      & \multirow{2}{*}{306}  & \multirow{2}{*}{$1.6\times 10^{12}$}  & $8\times 10^{10}$  & \multirow{2}{*}{Merger with an LMC-like companion}    \\
             & FIRE-2 CR-CD    &      &    &    & $4\times 10^{10}$  &     \\

        \midrule
        
        \multicolumn{7}{c}{\textbf{Dwarf galaxies}} \\
        \addlinespace

        \multirow{2}{*}{\texttt{m11c}} 	 & FIRE-2 MHD+  &   \multirow{4}{*}{$\approx 2\times 10^3$}   & \multirow{2}{*}{138}  & \multirow{2}{*}{$1.4\times 10^{11}$}  & $1\times 10^{8}$ & \multirow{2}{*}{Late-forming, LMC-like halo}    \\
             & FIRE-2 CR-CD    &      &    &    & $9\times 10^{8}$  &    \\
        \addlinespace
        \multirow{2}{*}{\texttt{m11b}} 	 & FIRE-2 MHD+ &   & \multirow{2}{*}{92.2}  & \multirow{2}{*}{$4.3\times 10^{10}$}  & \multirow{2}{*}{$8\times 10^{7}$} & \multirow{2}{*}{Disky, fast-rotating}     \\
             & FIRE-2 CR-CD    &      &    &    &   &    \\
        
        \bottomrule
        \end{tabular}}
    \caption{FIRE-2 simulation runs analyzed in this work. We choose two MW-mass haloes and two dwarf haloes. All simulations are zoom-in, and properties presented are at $z=0$. For more information about the haloes, see \citet{hopkins2018fire}.}
    \label{tab:sims}
\end{table*}

We adopt a standard, flat $\Lambda$CDM cosmology with $h\approx 0.70$, $\Omega_{\rm M}=1-\Omega_{\Lambda}=0.27$, and $\Omega_{\rm b}\approx 0.045$, consistent with the current constraint \citep{2014A&A...571A..16P}.

\subsection{B-fields and CRs}\label{sec:B-fields and CRs}

Both B-fields and CRs are included in the FIRE-2 simulations as part of the ``non-default'' physics, and can be found in the second FIRE-2 public data release \citep{wetzel2025second}. Here, we only recapitulate the essence of those implementations, and refer the readers to the original papers for more details: for the B-field physics, refer to \citet{hopkins2016accurate} for MHD implementation in \texttt{GIZMO} and to \citet{su2017feedback} and \citet{hopkins2020but} for MHD implementation in FIRE simulations; for the CR physics, refer to \citet{chan2019cosmic} and \citet{hopkins2021effects}. 

For the B-field, an initially homogeneous constant seed field was seeded at $z=100$ with magnitude $B_{{\rm ini,}x}=B_{{\rm ini,}y}=B_{{\rm ini,}z}=10^{-1}{\rm \, nG}=10^{-10} \, \Gauss$ in physical value, giving total $B_{\rm ini}=\sqrt{3}\times 10^{-10} \, \Gauss$ ($\sim 10^{-14}$ comoving G \footnote{Because of flux freezing, $B_{\rm comoving}=\frac{B_{\rm physical}}{(1+z)^2}$}.). The ideal MHD equations (i.e., no magnetic diffusion) were evolved. Anisotropic conduction and anisotropic viscosity were also included in the simulations. To mitigate the spurious divergence error (i.e. a non-zero $\nabla\cdot \vec{B}$), we employ a combination of Powell \citep{powell1999solution} and Dedner \citep{dedner2002hyperbolic} divergence cleaning schemes. The details of the methods and tests performed were presented in \citet{hopkins2016accurate}. 


As for the CRs, we employ a two-moment method that explicitly evolves the CR energy density $e_{\rm CR}$ and the CR energy flux $\vect{F}_{\rm CR}$. The most important distinction between the one-moment and two-moment methods is that the flux is calculated in the former but solved in the latter. The CR transport was introduced in the diffusion equation of CR flux. The CR diffusion coefficient (or diffusivity) $\kappa$ is a constant ($\approx 3\times 10^{29} {\rm \, cm^2 \, s^{-1}}$), whose value was constrained by matching empirical $\gamma$-ray observations from nearby galaxies \citep{chan2019cosmic}. The CR physics is coupled with the B-field, so we essentially evolve a set of CR-MHD equations. Note that in this work we only include the constant diffusion model with the aforementioned diffusivity value (dubbed ``CR-CD'' or ``CD'' hereafter), aiming to test if the presence of CR modifies the evolution of the B-field; discussion about other physically motivated models with variant diffusion coefficients, as well as their implications on galaxy properties and observables, can be found in \citet{hopkins2021testing} and \citet{lu2026constraining}.

\subsection{Selected simulations}\label{sec:selected sims}

In Table \ref{tab:sims} we introduce the simulation runs analyzed in this work. We choose a total of four target haloes, including two intermediate mass dwarfs (\texttt{m11b} and \texttt{m11c}) and two MWs (\texttt{m12i} and \texttt{m12f}). For each target halo, we ran the simulations with (labelled as ``CR-CD'') and without (labelled as ``MHD+'') the CRs. Main properties of the zoom-in galaxies at $z=0$ are also presented in Table \ref{tab:sims}. It should be noted that in general, the CR-CD run has smaller stellar mass compared to the MHD run for the same halo, and this difference is more pronounced as the halo becomes more massive. This is related to the fact that CRs provide non-thermal support and thus suppress star formation rates \citep{hopkins2020but, lu2026constraining}.

In this work we only present the aforementioned selection of simulation runs, though we remark here that other FIRE simulation runs of MWs and dwarfs exhibit similar B-field features, both the structure and evolutionary tracks, and therefore we state that this selection is representative enough to draw robust conclusions.

\section{B-field properties at $z=0$}
\label{sec:properties}

\begin{figure*}
    \centering
    \includegraphics[trim={0.0cm 0.0cm 0.0cm 0.0cm}, clip, width =0.98 \textwidth]{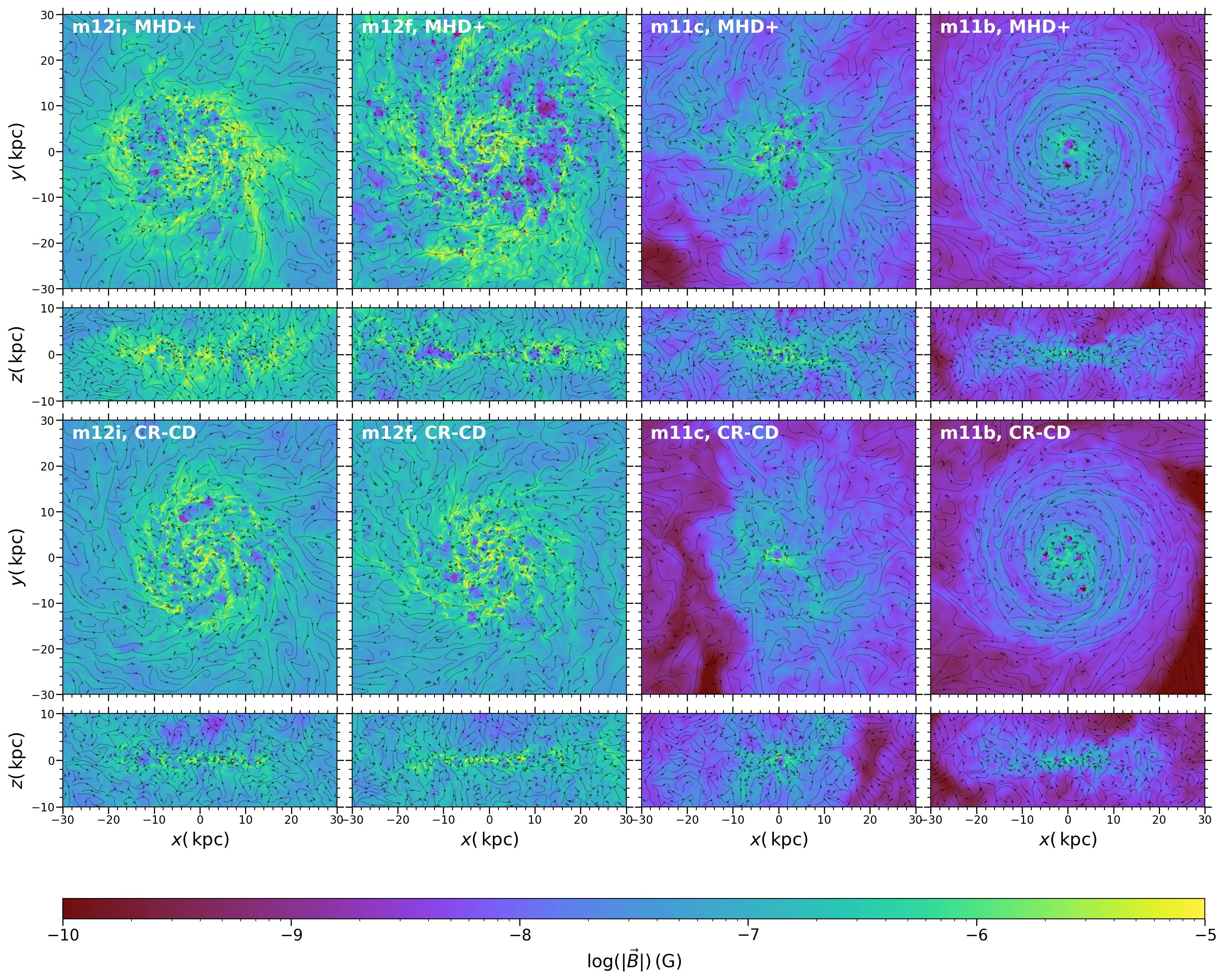}
    \caption{Visualization for the B-field in our selected galaxies at $z=0$. We show slice plots in both face-on (bigger panels) and edge-on (smaller panels) views. All slice plots are centred at the centres of the galactic disks. Overlaid streamlines illustrate the orientation of the B-field inside the sliced planes. ``Voids'' with low B-field strengths seen in the two MW runs correspond to SN explosions where the densities are low. The MWs (left two columns) have a much higher B-field strength compared to the dwarfs (the right two columns), especially outside the disk. The B-field is in general turbulent, but there is a slight alignment between the field lines and the galactic spirals due to rotations. The \texttt{m11b} has a fast-rotating disk, so the B-field lines outside the disk swirl around the disk.}
    \label{fig:B_fld_vis_with_streamlines}
\end{figure*}

In this section, we discuss properties of B-field in our simulated galaxies at $z=0$. A comprehensive analysis on structure and observable-inferred B-field properties for the $L_*$ galaxies has already been presented in \citet{ponnada2022magnetic, ponnada2023synchrotron}. Those works focused more on the distinctions between MHD and CR runs. Therefore, in this work, we focus more on the evolution of the B-field and the mass dependence of B-field properties.

In \fig{B_fld_vis_with_streamlines} we present a visualization for B-fields in our selected FIRE-2 galaxies at $z=0$. The colours represent B-field strengths, sliced through two planes representing the face-on and edge-on views of the galactic disks. Streamlines of B-fields are overlaid to illustrate the geometry of the B-fields. The morphology of B-field is in broad resemblance with the gas density distribution [not shown here, but see e.g. \citet{hopkins2020but}]: we see spiral-like structure of the B-field in the disk midplanes. The B-field lines are generally turbulent. Closer to the galactic disk, we see alignment between the field lines and the direction of the rotation of the disk. This feature is particularly pronounced in the \texttt{m11b} halo, as the B-field lines are almost swirling around the central fast-spinning disk. We see from our visualizations that the inclusion of CRs (i.e. MHD+ vs. CR-CD runs for the same halo) has little impact on the structure of B-field, while the B-fields features in different host haloes are more distinct. 

\begin{figure*}
    \centering
    \includegraphics[trim={0.0cm 0.0cm 0.0cm 0.0cm}, clip, width =0.98 \textwidth]{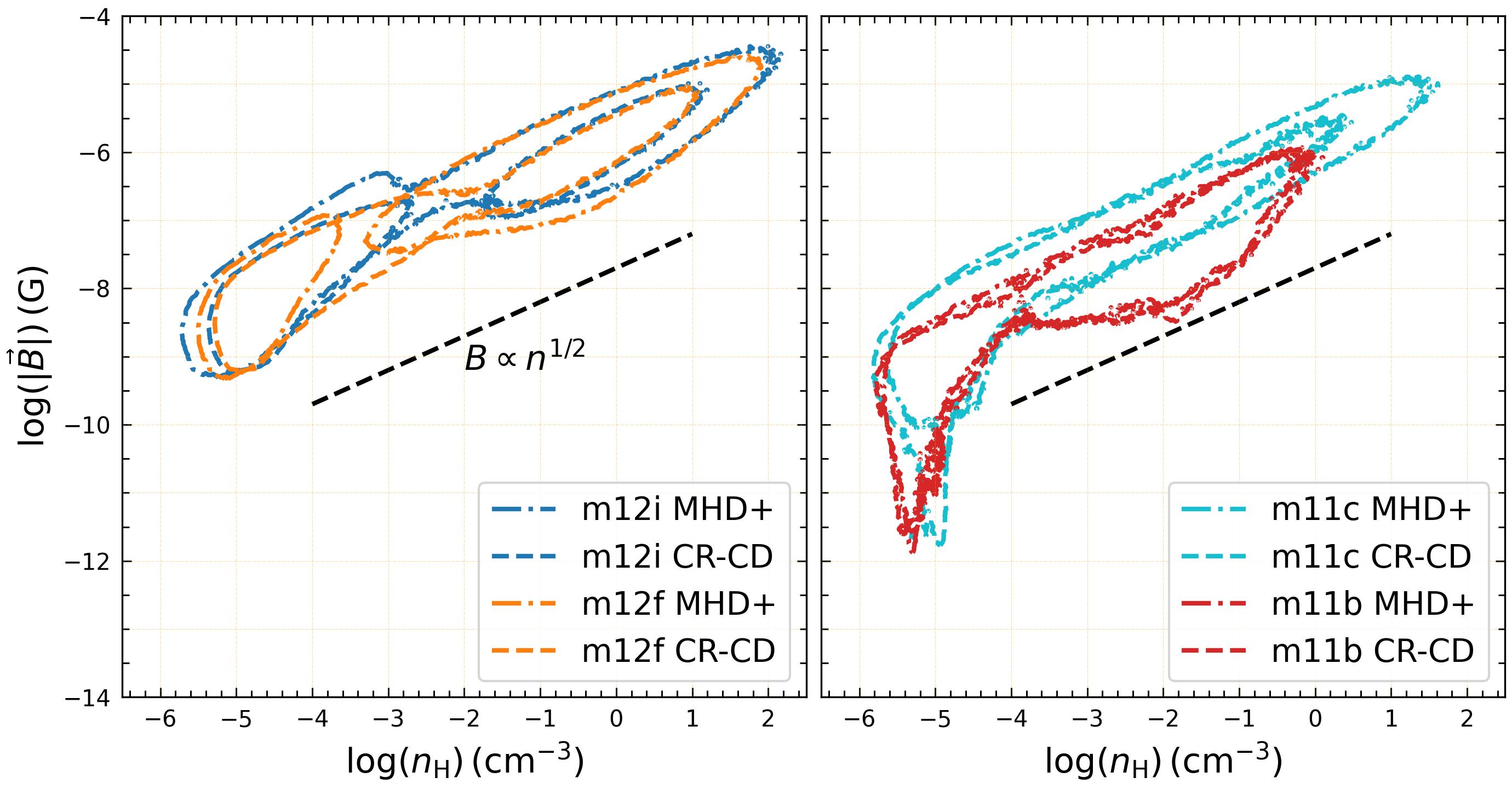}
    \caption{Mass-weighted $B-\nH$ distributions at $z=0$ within $R_{\rm vir}$. We present 90\%-enclosed highest-probability-region (HPR) contours of the 2D mass-weighted probability distribution function (PDF) in
    $(\log \nH,\,\log B)$ for the MW-mass (left) and dwarf (right) galaxies. Linestyles label the MHD+ and CR-CD runs for each halo. The dashed black line shows $B\propto n_{\rm H}^{1/2}$ for reference, which could be a direct result from assuming flux-freezing in a 2D planar or filamentary geometry (see the main text for a more detailed discussion). We see a general trend that the gas in the dwarfs have lower B-field strengths despite similar density ranges. This could be mainly due to that saturation is controlled by the turbulent energy (see discussion in \se{4.1}). 
    }
    \label{fig:B_nH_contours}
\end{figure*}

In \fig{B_nH_contours}, we show the distribution of the B-field strength $B=|\vec{B}|$ as a function of the density $\nH$. We first compute the 2D mass-weighted distribution histograms of $B(\nH)$ for all gas at $r\leq R_{\rm vir}$ for each run (where $r$ is the galactocentric distance), then draw contours enclosing $90\%$ of the highest-probability-region of such histograms. All the MW-mass galaxies are placed in the left panel and all the dwarfs are placed in the right panel. 
Both MW-mass galaxies and dwarfs roughly follow the $B\sim \nH^{1/2}$ in their $B-\nH$ distributions, with larger deviations from this trend at low densities. This could arise from various scenarios, and we discuss them below.
\begin{enumerate}
    \item Flux-freezing \citep{alfven1942existence}. This happens in ideal MHD and when the B-field is completely passive. In general, if the B-field collapses along with the environment, the B-field strength $B$ can be modelled by some power-law relation with the density $\nH$: $B\propto \nH^\alpha$. Such relations can arise from multiple scenarios depending mostly on the flow geometry \citep{tritsis2015magnetic}. The value of this parameter, in theory, can be any number (both positive and negative), but in reality the number is almost always between $0$ and $1$ due to geometric constraints (see \se{discussion} for more discussion). For an isotropic collapse or expansion (e.g. early expansion before structure formation), $\alpha\sim 2/3$; for 2D filamentary or planar collapses, which is a better representation for low-$z$ galaxy formation, $\alpha \sim 1/2$. The fact that $B\sim \nH^{1/2}$ for most of the gas within $R_{\rm vir}$ suggests that flux-freezing could be the dominant physical process at low-$z$. More detailed discussions about the $B-\nH$ relations can be found in \citet{tritsis2015magnetic, whitworth2025relation}, and we will also return to this in \se{discussion}.
    \item Dynamical equilibrium. This happens when the magnetic pressure gradient from the large-scale mean field in the disk becomes strong enough to resist the magnetic field line bending due to self-gravity: $|\nabla P_{\rm B}|\sim P_{\rm B}/R_{\rm disk}\sim g$, or $\frac{B^2}{R_{\rm disk}}\sim \rho \frac{v_{\rm circ}^2}{R_{\rm disk}}$. Given that the circular velocity $v_{\rm circ}$ is constant at a given radius, one could recover the $B\sim\rho^{1/2}$ relation\footnote{Although in this work we mostly present results regarding to the hydrogen density $\nH$, it is more or less equivalent to the total mass density $\rho$ in the galaxy context. We therefore do not make the distinction and use $\rho$ and $\nH$ interchangeably in the following discussion.}  from this setup. However, this applies only to very high density gas, primarily in the disk.
    \item Energy equilibrium. This happens when the saturation is resulted from tension shutting down the small-scale dynamos and magnetic helicity, then $P_{\rm mag}\sim P_{\rm turb}\sim \varepsilon \rho v_{\rm turb}^2$, namely the magnetic energy saturates to a certain fraction of the turbulent energy (which is what we observe in our simulations with $\varepsilon\sim 1-10\%$; see \fig{par_tracing_tracks}). From this we obtain $B\propto \sqrt{\rho}v_{\rm turb}$ (which is equivalent to $v_{\rm A}\propto v_{\rm turb}$). If $v_{\rm turb}$ is constant, which is plausible at high-density regime as argued in \citep{su2018stellar}, one again could recover the $B\sim\rho^{1/2}$ relation. 
\end{enumerate}
\no Since our simulations evolve ideal MHD, flux-freezing is the underlying continuum expectation and is likely a main driver of the trends we observe. However, finite-resolution numerical effects can introduce departures from exact flux-freezing, as discussed in \se{5.2}. On a much larger scale, the physical scenarios discussed above could have an impact on the scaling relations on top of flux-freezing, and we do not have a clear distinction here. Note that we do not apply rigorous fits to the results, and it is also hard to constrain the slopes of the power law observationally, since $1/2$ and $2/3$ are often not clearly separable \citep{crutcher2010magnetic, tritsis2015magnetic}.

\begin{figure}
    \centering
    \includegraphics[trim={0.0cm 0.0cm 0.0cm 0.0cm}, clip, width =0.49\textwidth]{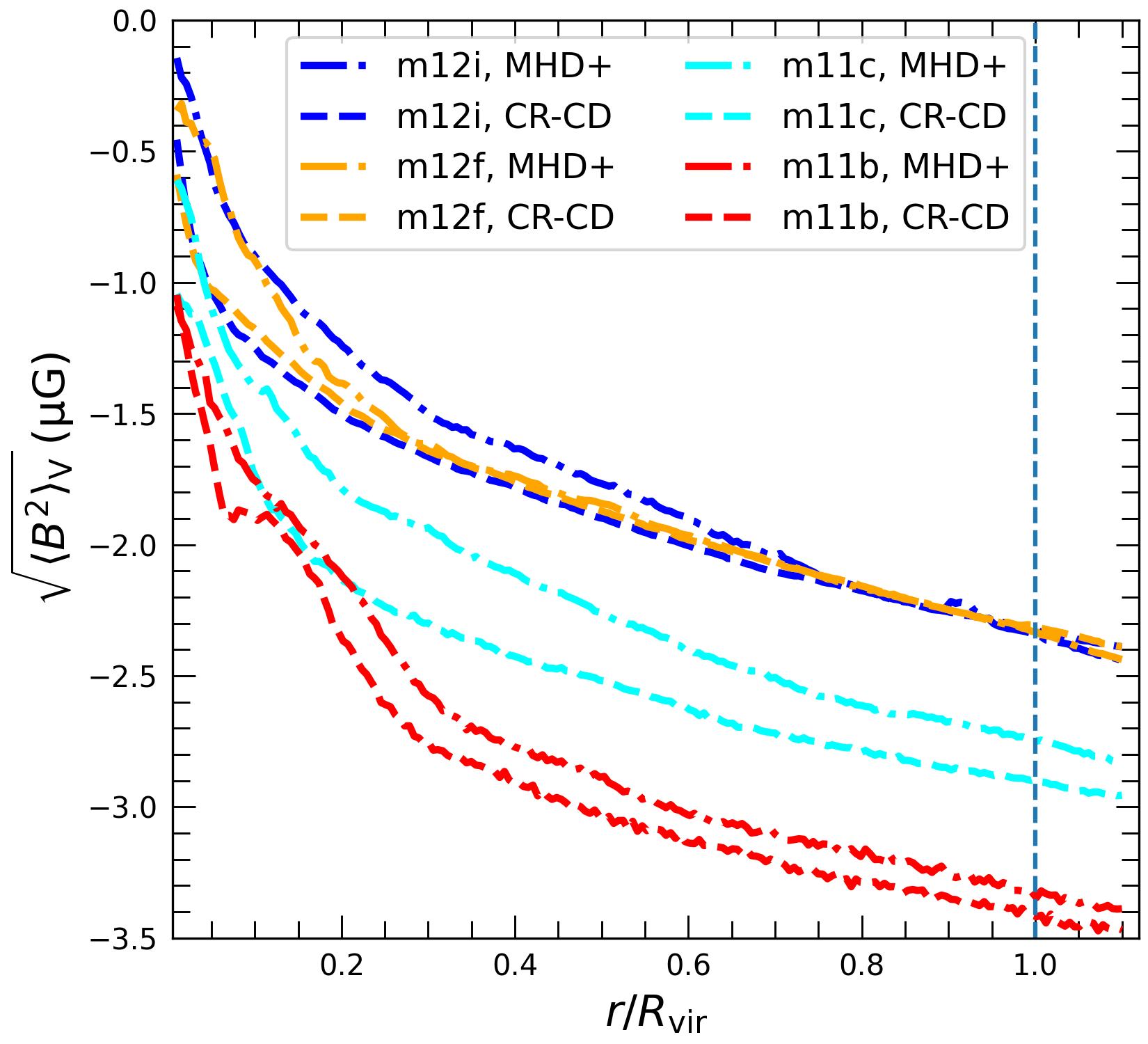}
    \caption{Radial profiles of the volume-weighted root-mean-square (rms) magnetic field strength $\sqrt{\left<B^2\right>_V}$ as a function of the galacto-centric radius $r$ for all our simulation runs at $z=0$. Each halo is colour coded in the same colour, and different linestyles correspond to the MHD and CR simulations for each halo, replicative to \fig{B_nH_contours}. Among different haloes, the B-field strengths differ more in the CGM than near the galactic centres.}
    \label{fig:B_prof}
\end{figure}

In \fig{B_prof}, we present the radial profiles of the volume-weighted root-mean-squared (rms) B-field strengths $B_{\rm rms}=\sqrt{\left<B^2\right>_V}$ in each of our simulation runs. The profiles are normalized to each halo's $R_{\rm vir}$. We again see a strong halo mass-dependence. Similar to what was presented in \fig{B_nH_contours}, the distinction between the two different MW-mass galaxies is much less significant than between the two dwarfs, and halo-to-halo variation is less pronounced than the mass dependency. All runs reach similar $\sim\mu\Gauss$-level B-field strength near the galactic centres, but begin to quickly diverge at $r\gsim 0.1 R_{\rm vir}$. These results are consistent with what was found in \citet{ponnada2022magnetic}.

We discuss the robustness of some of these basic B-field properties against resolution and feedback schemes later in \se{robustness}.



\section{The amplification of B-fields}
\label{sec:amplification}

In this section, we present how the B-field gets amplified over the course of galaxy formation. Here, we first present the results of our ``poster-child'' galaxy (\se{4.1}), the \texttt{m12i} CR-CD run to study its evolutionary course, then mention the results for the rest of the galaxies in our sample (\se{4.2}, full results in Appendix \ref{Misc plots}). 

\subsection{A case study of the \texttt{m12i} CR-CD run}\label{sec:4.1}

\begin{figure}
    \centering
    \includegraphics[trim={0.0cm 0.0cm 0.0cm 0.0cm}, clip, width =0.49\textwidth]{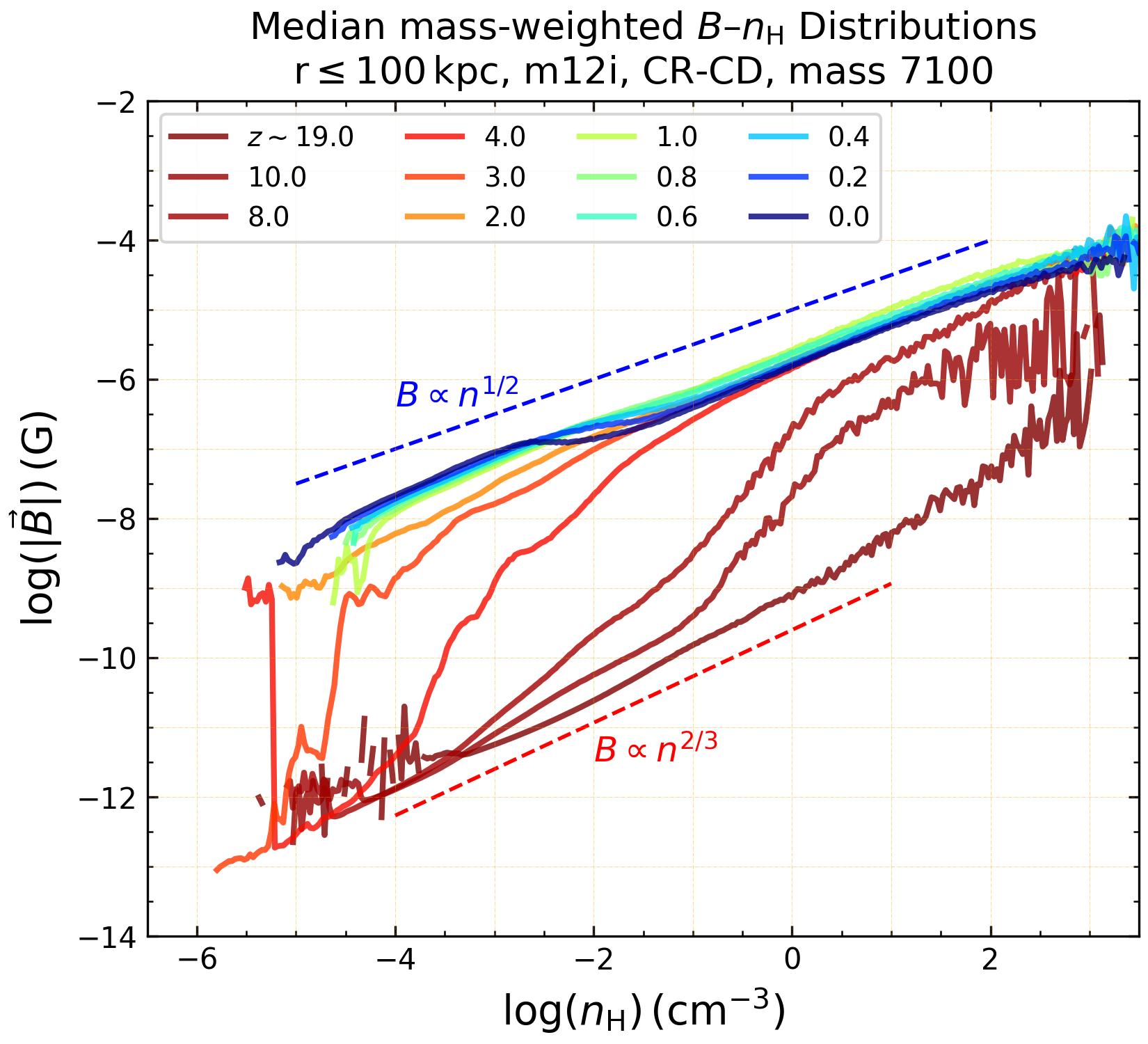}
    \caption{Distributions of $B$ over $\nH$ of the \texttt{m12i} CR-CD run at different redshifts. We compute the 2D $B-\nH$ mass-weighted PDFs for gas within $100 \kpc$ of each selected time, then take the mass-weighted median of $B$ over each $\nH$ bin. Colours correspond to redshifts/times, with the more reddish colours representing early Universe and the more bluish colours representing the current Universe. The $B\propto \nH^{1/2}$ and $B\propto \nH^{2/3}$ scaling relations are overlaid along with the median lines to guide the eye. Note that in the lower-left corner there appears to be gas with very low $\nH$ and $B$ in high-$z$. Those are gas particles out of the zoom-in regions.}
    \label{fig:B_nH_hist_evolution}
\end{figure}

In \fig{B_nH_hist_evolution} we show the statistical evolution of the B-field strength $B$. We choose a fixed spherical volume of $100\kpc$ (physical) radius and compute the 2D mass-weighted $B-\nH$ PDFs. For each resultant PDF, we take the mass-weighted median values of $B$ in each $\nH$ bin. Each line roughly depicts the ``ridgeline'' in the corresponding 2D phase histogram (if symmetric). We present results from $z\sim 19$ to $z\sim 0$ and colour-code the lines in redshifts. 

In the early Universe, $z\sim 19$, right after the initial conditions of our simulation, the B-field is weak and mostly follows the $B\propto \nH^{2/3}$ relation. This is because the early gas motion is mostly isotropic in 3D in the expanding universe and flux-freezing is the dominant process. As the redshift decreases, the B-field in the high-density gas gets amplified first, starting from $z\sim 10$ to $z\sim 4$. The B-field in the low-density gas gets amplified subsequently, starting around $z\sim 3$. After $z\sim 1$, the strength of the B-field stays roughly constant for the same density and follows the relation of $B\sim \nH^{1/2}$ (as already shown in \fig{B_nH_contours}). 
We assert that the B-field in the high-density gas gets amplified first because it is closer to the galaxy, and as soon as the structure formation kicks in and feedback becomes important (at $z\sim 10-4$), the turbulence and large-scale motion drives the amplification.

\begin{figure*}
    \centering
    \includegraphics[trim={0.0cm 0.0cm 0.0cm 0.0cm}, clip, width =0.98\textwidth]{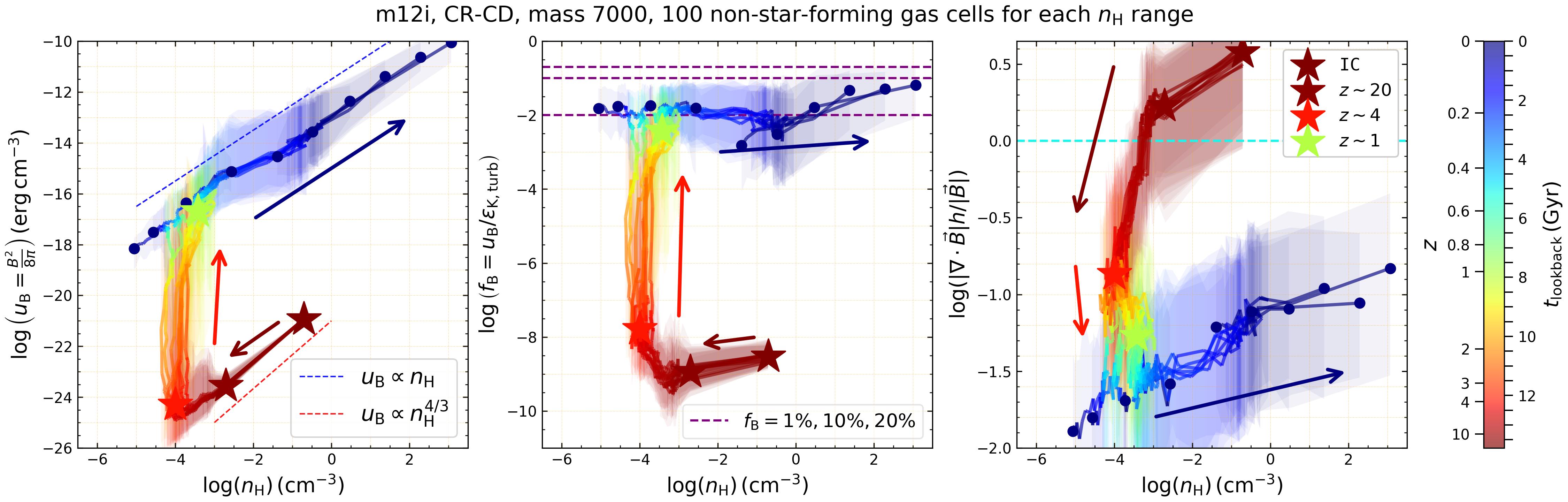}
    \caption{Tracks tracing individual gas particles for the \texttt{m12i} CR-CD run. The traced physical quantities are labelled at each panel's axes. We choose the particles with $\nH$ ranging from $10^{-6}-10^{3} \cmc$ at $z=0$ (so for a given $\nH$ profile, this essentially ensures including particles across different locations of the galaxy), and every individual track in each panel represents the median value of a group of 100 particles within the same density range at $z=0$ and is colour-coded by time and redshift. In each panel, arrows pointing from high to low redshifts are annotated to guide the eyes. The colour scheme transitions linearly in the look back time. Shaded regions under the tracks correspond to the 16th-84th percentile of the same group of particles at each time, colour-coded the same way as the tracks. See the main text for details of particle selection and grouping. The $z=0$ (final) locations for each group are marked as dots on the tracks. We label several significant evolutionary milestones using stars with the colours corresponding to the occurring times: the initial condition (\texttt{IC}) at $z\sim 99$, the first snapshot at $z\sim 20$, the structure formation kicking-in time $z\sim 4$, and the saturation time $z\sim 1$. With these milestones, a rough three-stage evolution of B-fields is observed. }
    \label{fig:par_tracing_tracks}
\end{figure*}

\begin{figure}
    \centering
    \includegraphics[trim={0.0cm 0.0cm 0.0cm 0.0cm}, clip, width =0.49\textwidth]{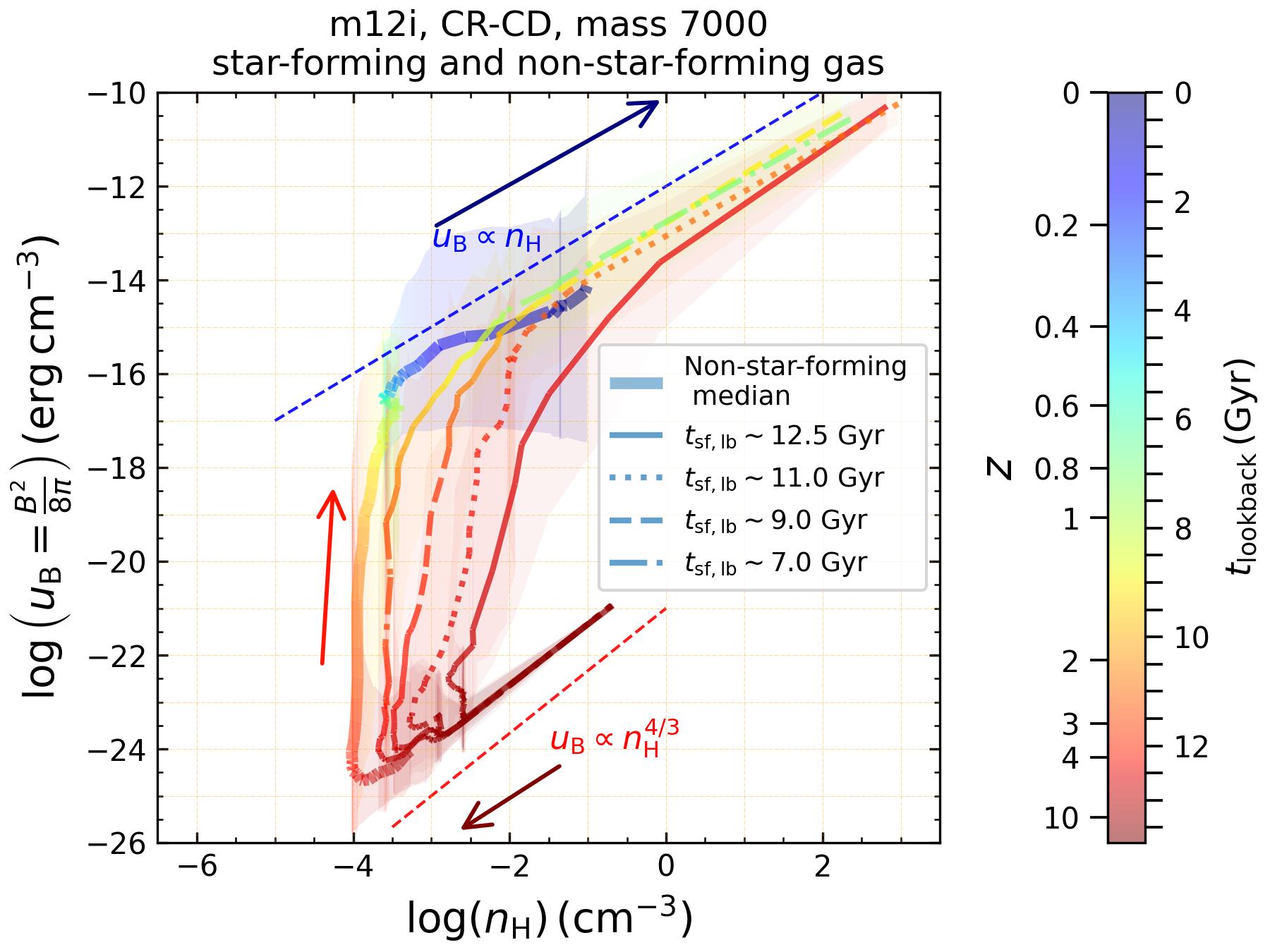}
    \caption{Tracks of $u_{\rm B}-\nH$, similar to the upper left panel of \fig{par_tracing_tracks}, but for star progenitors in the \texttt{m12i} CR-CD run. Guiding arrows are overlaid to guide the eye for the evolutionary orientation. We group the gas particles based on the times when they become stars (for $t_{\rm st, lb}\sim 12.5, 11, 9, 7 \Gyr$) and randomly choose 300 gas particles within each group. Each track now represents the median values for each star-forming group, labelled with different linestyles (see the legend). All tracks reach $\nH\sim 10^3 \cmc$ at the end, which is the star formation threshold. The median line of all particles chosen for \fig{par_tracing_tracks} representing all non-star-forming particles is also underlaid with a much thicker linestyle to guide the eye. The colour scheme is the same as in \fig{par_tracing_tracks}, and the shaded regions likewise correspond to the scatter among each group of particles at each time.}
    \label{fig:B_nH_tracks_star_progenitors}
\end{figure}

Note that what we show in \fig{B_nH_hist_evolution} is an Eulerian evolution, i.e., not tracing individual gas elements, and we do not know what causes the amplifications yet. To further investigate amplification, in \fig{par_tracing_tracks}, we employ a Lagrangian particle-tracing method to inspect the evolution of the B-field at the individual level of gas elements. We first select our particle samples by randomly choosing 100 particles in each of the following contiguous density ranges at $z=0$ with $r\leq R_{\rm vir}$: $\log\nH\in[-6, -5), [-5, -4), \dots , [2, 3), [3, 4)$. We then group the chosen particles based on their densities, so this gives us a total of 1000 gas particles (ten groups of 100 particles) chosen from $z=0$ \footnote{For simplicity, we exclude particles that have undergone particle splittings.}. We trace each gas element at each given snapshot and compute their density $\nH$, B-field energy density $u_{\rm B}=B^2/8\pi$, magnetic-to-turbulent ratio $f_{\rm B}\equiv u_{\rm B}/\varepsilon_{\rm K, turb}$ (where $\varepsilon_{\rm K, turb}$ is the filtered turbulent kinetic energy density; see \se{4.1.1} for its definition and further discussion), and the relative divergence error of B-field $|\nabla\cdot \vect{B}|h/|\vect{B}|$ [where $h=\left(3V/4\pi\right)^{1/3}$ is the effective size of the gas particle]. We compute the medians of those quantities for the 100 particles in each group at each snapshot and plot the tracks tracing their evolutions. The terminal location of each group's track is labelled by a dot. Note that in this exercise, by design, we only choose particles that have never formed stars. We present results for particles that do undergo star formation (or ``star progenitors'') later in \fig{B_nH_tracks_star_progenitors}.

Though from a different perspective, the Lagrangian particle tracing results paint a story reminiscent to our Eulerian results. First, we see small group-to-group variations among the tracks, which indicates that all particles undergo similar amplification processes, despite reaching different B-field strengths at $z=0$ due to the different densities. Several important milestones during the evolution are discernible along the tracks, and we flag them using stars in the figure. With these flags, we crudely categorize the amplification into the following three stages.

\begin{enumerate}
    \item \textbf{Initial expansion stage (from the initial condition to $z\sim 4$)}: this is the stage when the motion of gas is dominated by the expansion of the Universe. We see that almost all particles follow the $B\sim \nH^{2/3}$ (which gives $u_{\rm B}\propto B^2\sim \nH^{4/3}$) going from higher densities to lower densities with very small scatter. 
    
    \item \textbf{Fast amplification stage ($1\lsim z\lsim 4$)}: this is the stage when most of the rapid amplification of B-field for the gas currently in the MW-mass galaxy and its halo took place. We see that from $z\sim 4$ to $z\sim 1$ (corresponding to $t_{\rm lb}\sim 12.5 \Gyr$ to $t_{\rm lb}\sim 7.5 \Gyr$), the B-field strengths of all particles increase by $\sim 3-4$ orders of magnitude (corresponding to an increase of $\sim 6-8$ orders of magnitude in $u_{\rm B}$) almost without any change in density, while the total kinetic energy stays mostly constant with slight fluctuations (not presented in \fig{par_tracing_tracks}, but can be found in the right panel of \fig{bulk_and_turb_fields}). This results in a boost in the magnetic-to-turbulent ratio $f_{\rm B}$, going from $\sim 10^{-8}$ to $\sim 10^{-2}-10^{-1}$, reaching saturation. Galaxy evolution wise, this is a chaotic stage with violent inflows and outflows that could generate turbulence, and major merger events are prevalent \citep{kerevs2005galaxies, dekel2009cold, muratov2015gusty}. 
    
    \item \textbf{Final saturation stage ($z\lsim 1$)}: this is the stage when the B-field strength is mostly saturated and has a simple density dependence. The increase in $B$ is mostly due to the increase in density (i.e. flux-freezing). This is in the mirror mode of stage (i), but the gas is being ``concentrated'' rather than being ``diluted''. The scaling relation roughly follows $B\sim \nH^{1/2}$ (or $u_{\rm B}\sim \nH$). This likely corresponds to a combination of 2D flux-freezing, which could be either filamentary accretion or planar rotation) and saturation processes (see \se{properties}). At this stage, the galaxy has been formed and large scale gas inflow and outflow are less chaotic, but the turbulence is well-developed and controls the field strengths.
\end{enumerate}

During all three stages, the B-field has always been a \underline{passive} field, i.e. it evolves following the environmental evolution, as the magnetic-to-turbulent ratio $f_{\rm B}\equiv u_{\rm B}/\varepsilon_{\rm K, turb}$ is always small. Even during the rapid amplification stage, $f_{\rm B}\lsim 1\%$, and the saturation value is $\sim 1\%-10\%$, depending on the density. This number is relatively small compared to the results from \citet{ntormousi2020dynamo} and \citet{pakmor2024magnetic}, who found saturation value of $\sim 10- 20\%$, but is comparable to the results from \citet{federrath2014turbulent} and \citet{rieder2017small-II}, who found the saturation value is $\sim 2-5\%$ in idealized turbulent boxes and galaxy simulations. Analytic works such as \citet{rodrigues2015galactic} and \citet{schober2015saturation} both report a similar a few per cents magnetic-to-turbulent ratio. However, the aforementioned works are all different from ours to a certain extent, either with different set ups (e.g., meshless vs. mesh codes, idealized/cosmological/analytical) or different feedback recipes (e.g., FIRE vs. Auriga), and other factors like resolution, Mach number (i.e. supersonic vs. subsonic turbulence; see \citet{federrath2016magnetic}) could all affect this value, so the comparisons are not ``apples-to-apples'' \footnote{A thorough comparison between grid and particle code can be found in \citet{tricco2016comparison}. However, those simulations do have magnetic diffusion (non-ideal MHD).}. On the other hand, the definitions of turbulent fields are non-unique and could affect the results. We will discuss the decomposition of $u_{\rm B}$ and $u_{\rm K}$ into turbulent and bulk components and its implications on the amplification processes in \se{4.1.1}. It is also worth mentioning that not only the magnetic-to-turbulent ratios are different, the saturated B-field strengths in the CGM seen in FIRE simulations are also generally lower than other galaxy zoom-in simulation projects (e.g. Auriga) \citep{pakmor2020magnetizing, su2018stellar}, for largely unknown reasons.

Noticeably, during stage (iii), the tracks of magnetic-to-turbulent ratio (the middle panel of \fig{par_tracing_tracks}) appear to first ``dip down'' then ''rise up'', after reaching $\nH\sim 1 \cmc$. This is due to similar ``dip-then-rise'' in both $u_{\rm B}$ and $\varepsilon_{\rm K, turb}$ tracks\footnote{This feature for $u_{\rm B}$ can be clearly seen in the left panel of \fig{par_tracing_tracks}. We do not present medians of $\varepsilon_{\rm K, turb}$ for each density group, but the individual tracks and the overall density dependence of $\varepsilon_{\rm K, turb}$ are presented in \fig{bulk_and_turb_fields}. A similar trend is visible there.}, but the former undergoes a more significant transition. 
We note that in the left panel of \fig{par_tracing_tracks}, particles actually fit better to the $B\propto \nH^{1/2}$ line after this ``dip'', and this could be related to gas elements joining the disk to exhibit a more planar geometry. This is still consistent with flux-freezing, but the properties of turbulence changes when in the disk compared to the CGM \citep{stern2024accretion}, so the turbulence growth is not as strong as the B-field growth.
This could also be related to clumpy medium (e.g. GMCs) near the galactic centre. We note that this is actually a robust trend that appears in every selected simulation (see Appendix \ref{Misc plots}).

The right most panel of \fig{par_tracing_tracks} presents the median relative divergence error of our traced particles. We see that the value of this parameter is relatively large ($\sim 1$) during the early times but drops to $\lsim 10\%$ after $z\lsim 4$, when the amplification stage started. The value only fluctuates and is always $\lsim 20-30\%$ at $z\lsim 1$. However, the quantity we present here could only indicate the relative dynamical importance of the residual divergence error, and a small dynamical contribution relative to other force terms in the MHD momentum equation does not necessarily mean no impact on amplification. We will discuss this in more detail in \se{discussion}.


As mentioned earlier, our selection method excludes all star-forming gas particles by design, which could bias our results. Therefore, in \fig{B_nH_tracks_star_progenitors}, we present the $u_{\rm B}-\nH$ tracks for star-forming particles, grouped based on their star formation times. We choose the following four look back time values $t_{\rm look back}\sim 7, 9, 11, {\rm and\,} 12.5\Gyr$ as our target times, and randomly choose 300 gas particles (``star progenitors'') that formed stars with $\pm 10 \Myr$ around our target times. Overall, we see that star progenitors undergo a similar amplification process compared to non-star forming gas. The difference lies in the timing: the star progenitors undergo the rapid amplification stage much earlier than the non-star-forming particles, because these gas elements have collapsed into the galaxy and been exposed to turbulent amplification earlier. This further confirms our argument on the passivity of the B-field and complements our earlier results: e.g., we see that the high-density gas starts getting amplified as early as $z\sim 10-5$ in \fig{B_nH_hist_evolution} but not in \fig{par_tracing_tracks}, since \fig{par_tracing_tracks} does not include early collapsing gas that already turned into stars. Therefore, the aforementioned three-stage evolution is universal and is regardless of gas density, though the timings shift for star progenitors. For galaxies with much smaller masses (e.g. ultra faint dwarfs), the timings could also shift due to a similar reason.

We experimented with different sets of randomly selected particles and all aforementioned results converged (not shown).

\subsubsection{Decomposition of the magnetic and kinetic fields into bulk and turbulent components using a Gaussian filter}\label{sec:4.1.1}

\begin{figure*}
    \centering
    \includegraphics[trim={0.0cm 0.0cm 0.0cm 0.0cm}, clip, width =0.98\textwidth]{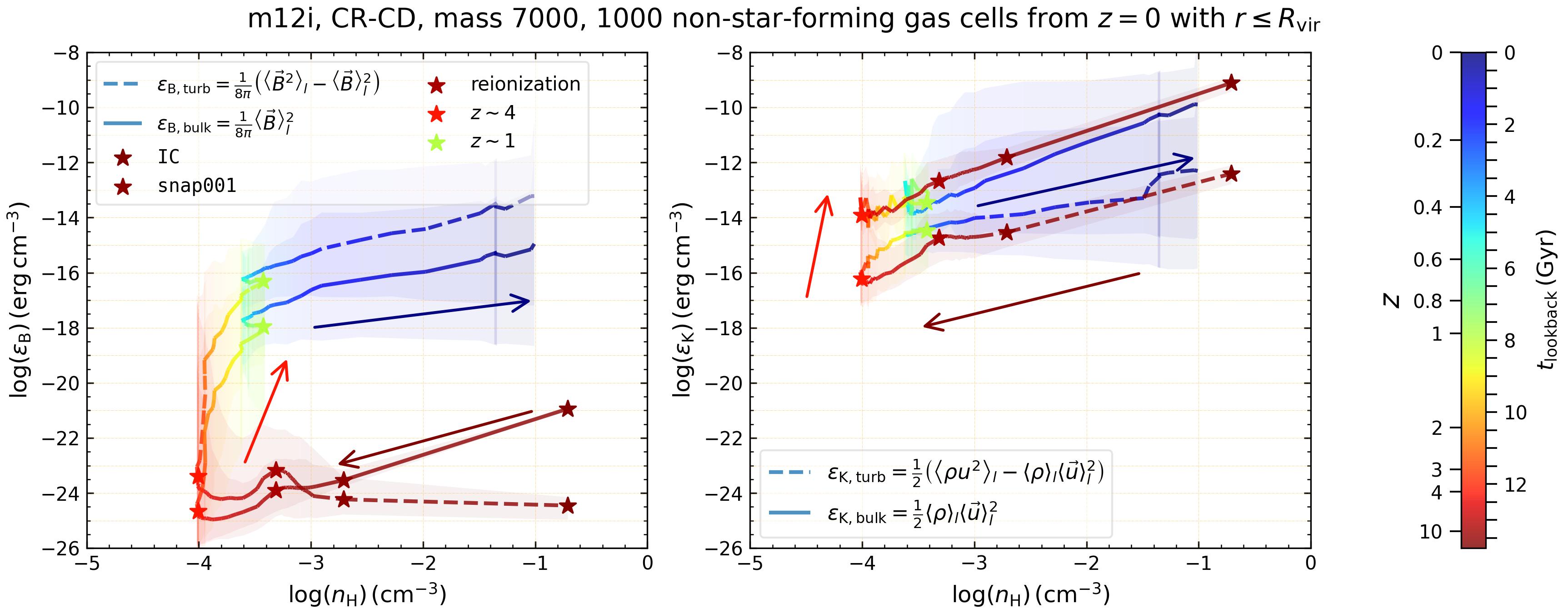}
    \caption{\texttt{m12i} CR-CD run's particle tracing tracks for the decomposed magnetic (left panel) and kinetic (right panel) energies, including the bulk (solid lines) and turbulent (dashed lines) components. The presented tracks trace the same set of particles in \fig{par_tracing_tracks}. The axis ranges are set to be the same on purpose to illustrate the dominance of kinetic energy throughout the galaxy evolution. Here, we adapt a particle-based filtering size $l=h$ where $h$ is the SPH smoothing length. We decompose the fields according to \equs{e_B_bulk}-\equm{e_K_turb} of each individual traced particle (the same ones presented in \fig{par_tracing_tracks}) and take the median values of the decomposed fields of all 1000 particles at each snapshot. Shaded regions with the same colour scheme again correspond to the 16th-84th percentile for each group of particles at each time. Guiding arrows are again overlaid to indicate the evolutionary track orientation. In addition to the original 4 milestones in the three-stage amplification process, we also label the rough time for the reionization, which is potentially responsible for the two humps seen in the tracks of $\varepsilon_{\rm B, turb}$ and $\varepsilon_{\rm K, turb}$ at $z\sim 9-10$.}
    \label{fig:bulk_and_turb_fields}
\end{figure*}

As mentioned earlier in this paper as well as discussed extensively in other literature, turbulent dynamos have been thought to be the main driver of B-field amplification, potentially responsible for the steep jump in $u_{\rm B}$ seen in \figs{par_tracing_tracks} and \figss{B_nH_tracks_star_progenitors}. To understand this, one needs to decompose the kinetic energy into bulk and turbulent components and compare to the unordered, turbulent B-fields. Here, we follow a filtering method introduced in \citet{hollins2022mean} and later used to study the intracluster medium (ICM) in  \citet{perrone2026characterizing}. We employ this method to decompose the magnetic and kinetic energies into bulk and turbulent components, but we stress that the decomposition methods are not unique. In the following, we briefly discuss the definitions of our bulk and turbulent fields and refer the readers to the original papers for a more thorough study.

We employ a three dimensional Gaussian filter with a window length $l$ to filter the fields. The filter function is defined as
\be
    W_l(\vec{x}, \vec{x}')=\frac{1}{(2\pi l^2)^{3/2}}\exp\left(-\frac{1}{2}\frac{|\vec{x}-\vec{x}'|^2}{l^2}\right).
\ee
One can verify that it satisfies the normalization condition. Then, for any given generic field (a scalar field or a component of a vector field) $f(\vec{x})$, the filtered field is defined as
\be
    \left<f\right>_l(\vec{x})=\int d^3\vec{x}'f(\vec{x}')W_l(\vec{x}, \vec{x}').
\ee
For discretized gas elements in our simulations, the above filtering is performed by summing over quadratures. From this we can define the bulk and turbulent components of the magnetic energy as

\be\label{eq:e_B_bulk}
    \varepsilon_{\rm B, bulk}\equiv \frac{1}{8\pi}\left<\vec{B}\right>_l^2
\ee
and
\be\label{eq:e_B_turb}
    \varepsilon_{\rm B, turb}\equiv \frac{1}{8\pi}\left(\left<B^2\right>_l - \left<\vec{B}\right>_l^2\right).
\ee
Similarly, we define the bulk and turbulent \footnote{Some other works (e.g. \citet{perrone2026characterizing}) further decompose the turbulent term into a ``cross term'' , $\varepsilon_{\rm K, cross}=\sum_i \left[\left<v_i\right>_l\left(\left<\rho v_i\right>_i-\left<\rho\right>_l\left<v_i\right>_l\right)\right]$, and a residual term (and call it the ``turbulent'' term). However, we do not make the distinction in our case as we found that the cross term is completely negligible (not shown) compared to the total turbulent term and has no strong physical significance.} components of the kinetic energy as

\be\label{eq:e_K_bulk}
    \varepsilon_{\rm K, bulk}=\frac{1}{2}\left<\rho\right>_l\left<\vec{u}\right>_l^2
\ee
and 
\be\label{eq:e_K_turb}
    \varepsilon_{\rm K, turb}=\frac{1}{2}\left(\left<\rho u^2\right>_l - \left<\rho\right>_l\left<\vec{u}\right>_l^2\right).
\ee
Note that here, $\vec{u}$ is the total physical velocity of the gas, not the peculiar velocity $\vec{v}$ that is explicitly evolved in our simulations. The difference is the Hubble flow:
\be
    \vec{v}=\vec{u}-H(a)\vec{r},
\ee
where $H(a)\equiv \dot{a}/a$ is the Hubble parameter. There have been debates on whether or not one should include the Hubble flow when calculating the turbulent kinetic energy. In this exercise, we follow \citet{perrone2026characterizing}, who included the Hubble flow term. Adding the Hubble flow back contributes little to the turbulent component, but will bump the bulk component, especially at high-$z$, when the Hubble expansion rate was faster.

Note that locally, $\varepsilon_{\rm B, bulk}+\varepsilon_{\rm B, turb}\not= u_{\rm B}$ and $\varepsilon_{\rm K, bulk}+\varepsilon_{\rm K, turb}\not= u_{\rm K}$, but when integrated over a large volume, the integrated energies will converge to the total energies. Here, we are not interested in recovering the total energies; rather, we seek correlations between the turbulent components of both. See \citet{perrone2026characterizing} for a more detailed discussion on this issue.

In \fig{bulk_and_turb_fields} we present the results of this decomposition, in a format similar to \fig{par_tracing_tracks} and \ref{fig:B_nH_tracks_star_progenitors}. We trace the same sets of particles from \fig{par_tracing_tracks}, except that we drop the groupings and present a single track for each quantity from the median values \footnote{Grouped tracks for each density group, without a surprise, mostly overlap each other except extending to different final densities (similar to \fig{par_tracing_tracks}).}. We apply the smoothing globally first then decompose the energies for individual gas elements. We take the smoothing length $h$ of each individual gas particle as the filter size, which ranges from $\lsim 0.5 \kpc$ to $\sim 10 \kpc$ depending on the density for this presented exercise, though the results are robust with respect to different filtering sizes, including for both $l\propto h$ and fixed $l$ (see Appendix. \ref{Misc plots}). We see that the bulk kinetic energy always dominates over the turbulent kinetic energy, but for the magnetic field, the bulk magnetic energy only dominates at very early time.

The turbulent magnetic energy experiences a sudden bump at $z\sim 9-10$ (at the lower-left corner on the track), which is also observed (but less pronouncedly) for the turbulent kinetic energy. We suspect that this is related to the start of the meta-galactic UV background implemented in our simulations \citep{faucher2009new} that suddenly heats the gas and creates strong pressure disequilibrium. This potentially leads to turbulent motion of ionized gas that can then magnify the B-field \footnote{The fact that the bump does not persist and both the turbulent kinetic and magnetic energies drop back suggests that this is not due to earlier stellar feedback.}.
Recent detailed simulations that model ionization fronts and the reionization process show that kpc-scale turbulence develops during this process, which creates pressure changes and induces instabilities and local turbulence \citep{cain2026kiloparsec, mcdermott2026weibel}. While we do not model ionization fronts in our simulations and do not resolve all relevant scales, we suspect that the onset of the global UV background temporarily creates turbulent motion in our simulations, partially mimicking this process.
A study on how the reionization drives turbulence and further amplifies the magnetic fields in the early Universe is beyond the scope of this work.


After the bump around $z\sim 9-10$, all tracks evolve in broad resemblance to \fig{par_tracing_tracks}. The magnetic field becomes dominated by the turbulent component later on, while the turbulent kinetic field is also being amplified at the same time. This suggests a correlation between the turbulence and the turbulent magnetic fields.
We discuss more on this in \se{discussion}.

\begin{figure*}
    \centering
    \includegraphics[trim={0.0cm 0.0cm 0.0cm 0.0cm}, clip, width =0.998\textwidth]{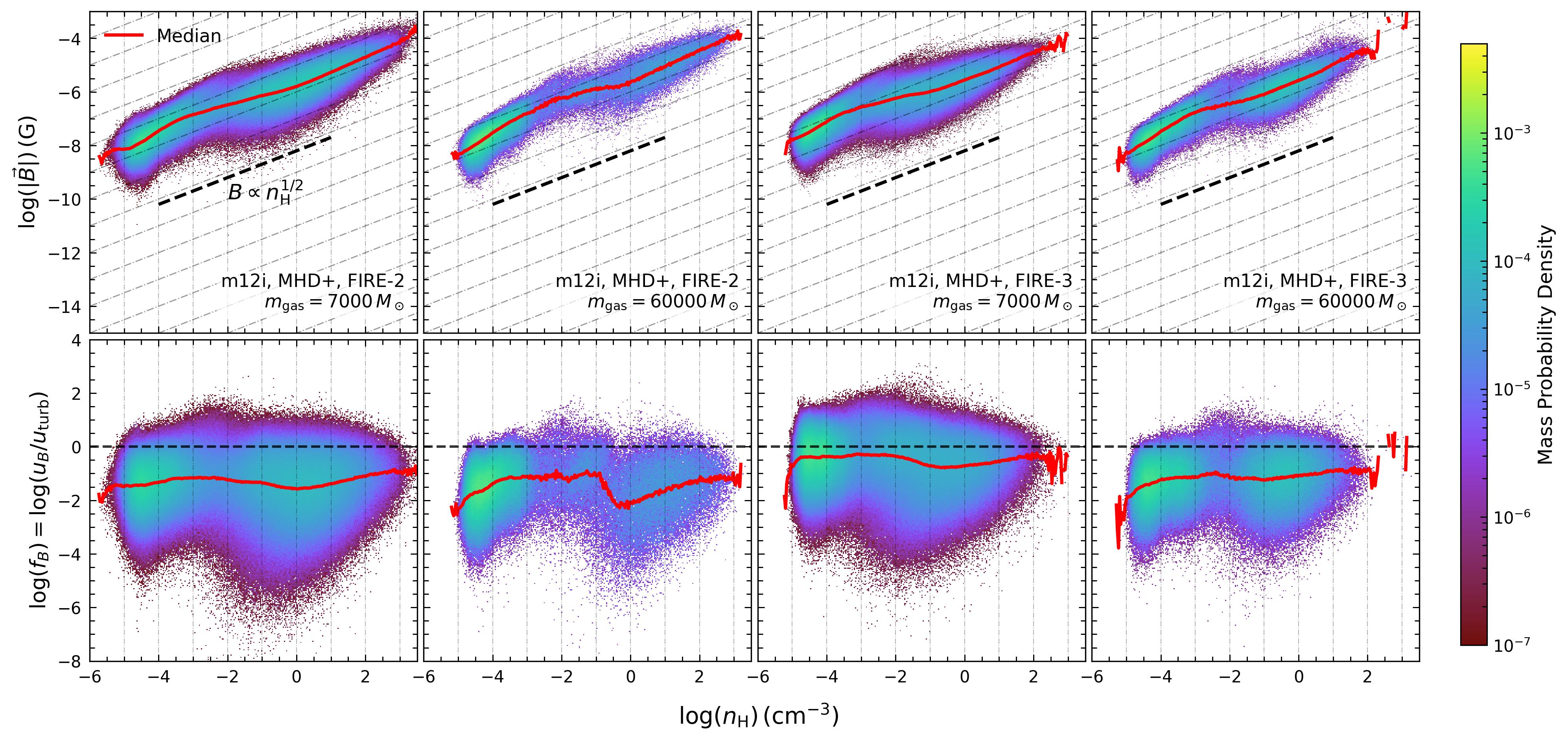}
    \caption{Mass-weighted $B-\nH$ (top panels) and $f_{\rm B}-\nH$ (bottom panels) distribution histograms at $z\sim 0$ for gas within $r\leq 150\kpc$ (see \se{4.1.1} for the definition of $f_{\rm B}$). We show the MHD runs for the \texttt{m12i} galaxy, with different resolutions and feedback schemes (FIRE-2 vs. FIRE-3). 
    The top and bottom panels of each column show the results for the same run (labelled at the lower-right corners of the top panels). For each top panel, we overlay the same set of $B\propto \nH^{1/2}$ lines, each separated by $1$ dex vertically, to guide the eye for comparisons of $B$ at the same densities. For each bottom panel, we put the equipartition line of $f_{\rm B}=1$ (corresponding to $u_{\rm B}=\varepsilon_{\rm K, turb}$) for reference. We see a broad resemblance amongst all variants for their $z\sim 0$ B-field properties. Slight differences in $f_{\rm B}$ appear between FIRE-2 and FIRE-3 runs (see the main text for our speculation), though an overall robustness against both the resolution and feedback schemes is confirmed.}
    \label{fig:robustness_B_fB_nH}
\end{figure*}

\subsection{Other selected simulations}\label{sec:4.2}
We ran the same particle tracing pipeline for the other runs from \tab{sims} and found similar trends for all the aforementioned results, including other MW-mass galaxies, dwarfs with and without clear disks. In particular, the B-field strengths (or energy densities) undergo similar three-stage evolutions. The $B\sim \nH^{2/3}$ and $B\sim \nH^{1/2}$ trends are still observed in the early and later evolutionary stages, though the saturated strengths depend on the halo mass. The magnetic-to-turbulent ratio $f_{\rm B}=u_{\rm B}/\varepsilon_{\rm K, turb}$ is consistent amongst different haloes, all reaching $\sim 1\%$ for $z\lsim 1$, so the saturated B-field strength is similarly controlled by turbulence. A similar dip-then-rise feature of $f_{\rm B}$ is seen in all runs. Similar humps of the turbulent magnetic and kinetic energies are seen around the epoch of reionization, but both go down and the major amplification is still driven by the formation of galaxy, feedback, and turbulence after $z\sim 4$. We present these results in Appendix \ref{Misc plots}.

\section{Discussions}
\label{sec:discussion}

\subsection{Robustness tests on resolution and feedback schemes}\label{sec:robustness}

In \ses{properties} and \ref{sec:amplification}, we focused on the B-field properties of the high-resolution FIRE-2 runs (e.g. $m_{\rm gas}\approx 7000,\Msun$ for MW-mass galaxies). 
In this section, we test the robustness of some of the basic results against numerical resolution and feedback schemes.

In \fig{robustness_B_fB_nH}, we present the mass-weighted distribution histograms of both the magnetic field strength $B$ and the magnetic-to-turbulent ratio $f_{\rm B}\equiv u_{\rm B}/\varepsilon_{\rm K, turb}$ as a function of hydrogen number density, $\nH$, where $u_{\rm B}=B^2/8\pi$ and $\varepsilon_{\rm turb, K}$ is the turbulent (non-bulk) component of the filtered kinetic energy density $\frac{1}{2}\rho u^2$, representing the strength of the local random motion of the gas element. See \se{4.1.1} for an exact definition [\equ{e_K_turb}] and a more detailed discussion of what this fraction means physically.


We compare our high-res ($m_{\rm gas}\approx 7000\Msun$) FIRE-2 \texttt{m12i} MHD run with its low-res ($m_{\rm gas}\approx 60000\Msun$) FIRE-2, high-res FIRE-3, and low-res FIRE-3 variants. FIRE-3 feedback differs from FIRE-2 primarily in the details of local coupling of the momentum and energy from SNe. The effect of this change results in \texttt{m12} galaxies in FIRE-3 forming faster during early stages but ending up with lower late-time SFRs and stellar masses \citep{hopkins2023fire}. Note that the FIRE-3 runs we include here contain no active BHs, and therefore no AGN feedback.

There is little difference in the median B-field strength at a given $\nH$ among all runs \footnote{The low-res runs contain fewer particles within $r\leq 150\kpc$, so the distributions appear to be sparser and more concentrated.}. The FIRE-3 high-res run's median $|B|$ is slightly higher than the other runs, by $0.3\sim 0.5$ dex (so $u_{\rm B}=B^2/8\pi$ will be higher by less than $1$ dex). However, the differences are more pronounced between the high-res FIRE-3 run and the rest of the simulations in the $f_{\rm B}-\nH$ distributions: its $f_{\rm B}$ values are higher than the other runs by $\sim $1 dex. Given the more similar magnetic energy densities, the bulk of the difference arises from differences in turbulent energy densities. It is not entirely clear why the B-field in the FIRE-3 run saturates at a higher fraction of the turbulent energy: although lower late-time SFRs in FIRE-3 could cause a lower level of turbulent motion, the difference in the $f_{\rm B}$ actually develops right after the amplification at $z\sim 1-2$ and high-res FIRE-3 simulations complete fast amplification at a similar time as high-res FIRE-2 (now shown), making it less likely that differences in SFRs at late times are the main cause. 
The detailed mechanism causing this difference will be studied in future work. 

In Appendix \ref{Misc plots}, we include an example of another simulation run with both versions of the FIRE feedback schemes at an earlier time, showing a similar level of differences in $f_{\rm B}$.


Overall, we find that the stages of the evolution and general properties of the B-field at $z=0$ are robust against both the resolution and the feedback schemes. 
However, the amount of turbulent energy going into saturation of the B-fields may vary for different numerical implementations. We leave a more detailed study of numerical effects on B-field evolution for future work.

\subsection{Origin of the rapid growth of B-fields}\label{sec:5.2}

In \se{4.1} we see a rapid growth stage of B-fields from $z\sim 4-1$ (\figs{par_tracing_tracks} and \figss{B_nH_tracks_star_progenitors}) with both the bulk and turbulent components being rapidly amplified (\fig{bulk_and_turb_fields}), and evolutions deviated largely from any possible power-law relations between $B$ and $\rho$. 
Without this stage, the B-field would only go up and down along pre-determined tracks, unable to attain the $\mu\Gauss$-level field at $z\sim 0$ seen in our simulations. We have claimed that this is potentially associated with the dynamos. In this section, 
we first discuss the theoretical possibility of such surges, then attempt to explain the reason for such growth in our simulations. We approach with both physical and numerical insights and propose future study plans to further investigate this issue.

To begin, we first review the flux-freezing law and its implications. In ideal MHD, the B-field and the flow co-evolve following the Faraday's induction equation:
\be\label{eq:induction}
    \pdv{\vec{B}}{t}=\nabla\times(\vec{v}\times\vec{B}),
\ee
assuming zero magnetic diffusivity $\eta_{\rm B}=0$. An important implication from this is the conservation of magnetic flux $\Phi_{\rm B}$ through a comoving Lagrangian surface -- the Alfven's theorem \citep{alfven1942existence}. However, it is not trivial to trace a comoving surface in simulations run with meshless codes like \texttt{GIZMO}, so a more practical, equivalent result that one could derive from \equ{induction} is the Lagrangian equation of evolution for the field strength $B$:

\be\label{eq:logB evolution}
    \frac{d\ln B}{dt}=\vec{b}^T(\nabla\vec{v})\vec{b}-\nabla\cdot\vec{v},
\ee
where $\vec{b}=\vec{B}/B$ is the directional vector along the B-field lines. Note that \equ{induction} and \equ{logB evolution} do not translate reciprocally, as to reach the latter from the former, one needs to explicitly assume $\nabla\cdot\vec{B}=0$, which raises complexities in MHD galaxy simulations because zero divergence is not guaranteed in a violent environment like FIRE galaxies even with cleaning schemes. Note that the right hand side of \equ{logB evolution} depends solely on the kinetic structure of the flow and the B-field, so we call it $\left(\frac{d\ln B}{dt}\right)_{\rm kin}$ hereafter. 

To build a connection to $B-\rho$ scaling relations, we substitute the continuity equation

\be
    \frac{d\ln \rho}{dt}=-\nabla\cdot\vec{v}
\ee

\no into \equ{logB evolution} to get

\be\label{eq:logB-logrho}
    \frac{d\ln B}{dt}=\left[1-\frac{\vec{b}^T(\nabla\vec{v})\vec{b}}{\nabla\cdot\vec{v}}\right]\frac{d\ln \rho}{dt}.
\ee

\no Letting 

\be\label{eq:alpha_eff}
    \alpha_{\rm eff} \equiv 1-\frac{\vec{b}^T(\nabla\vec{v})\vec{b}}{\nabla\cdot\vec{v}}=-\frac{\left(\frac{d\ln B}{dt}\right)_{\rm kin}}{ \nabla\cdot\vec{v}},
\ee

\no one could recover the $B-\rho$ scaling relation from \equ{logB-logrho}: 

\be
  \frac{d\ln B}{d\ln \rho}=\alpha_{\rm eff} \quad {\rm or} \quad B\propto \rho^{\alpha_{\rm eff}}. 
\ee

\no This corresponds to the power-law relations that we see during various stages of the evolution of the B-field (\figs{B_nH_contours}, \figss{B_nH_hist_evolution}, and \figss{par_tracing_tracks}). Note that physically $
\alpha_{\rm eff}$ [or equivalently $\left(\frac{d\ln B}{dt}\right)_{\rm kin}$] represents how much the flow strain tensor is projected along the B-field lines, which depends solely on the flow and the geometry of the B-field. For example, a negative $\nabla\cdot\vec{v}$ represents local density increase, which indicates a positive $\alpha_{\rm eff}$ that will result in B-field amplification. 

There are a few subtleties that need to be emphasized here. First, \equ{alpha_eff} only depicts the microscopic evolution of B-field, essentially indicating that the solenoidal mode (with a diminutive $\nabla\cdot\vec{v}$) is the dominant mode for the amplification of the B-field \citep{federrath2016magnetic}, which could result in a large $|\alpha_{\rm eff}|$. However, such amplification modes powered by the conventional ``stretching, twisting, and folding'' of field lines [i.e. $\alpha-\Omega$ dynamos, powered by magnetic/current and kinetic helicities; see \citet{ntormousi2020dynamo} and \citep{federrath2016magnetic}] occur on scales below the resolution limit of our simulations, so simply tracing individual particles as we did in \se{4.1} would not reflect it. 

Second, it is a common practice to discuss dynamos (particular large scale dynamos) in terms of the Reynolds-averaged quantities [e.g., $\left<B^2\right>$ vs. $\left<B\right>^2$, see \citet{2017MNRAS.464L.105E, brandenburg2023galactic}], and a lack of amplification in small scale $B$ does not necessarily warrant the same for large scale $B^2$, as the spectrum of magnetic field is highly scale-dependent \citep{kulsrud1992spectrum}. The interpretation of \equ{alpha_eff} could vary largely depending one's definition of a dynamo, and though topological rearrangement and sustaining complete dynamos in galaxy usually require magnetic diffusion, an amplification on large scale is still possible in ideal MHD. 

The above discussion is purely theoretical. In our simulations, we see a close-to-infinity slope in the $\log u_{\rm B} - \log \nH$ tracks (\fig{par_tracing_tracks}). Although such surges might be explained by the dynamo arguments established in the above discussion, these physical processes happen on scales much smaller than our simulation resolution limit. We therefore approach with a different perspective -- we propose that amplification is initiated by numerical effects, which have been shown to be important to B-field evolution in galaxy simulations \citep{wissing2023numerical, pakmor2026comparison}. 
It is likely that the role of diffusion (i.e. non-ideal MHD) to drive such strong amplifications, once the local turbulence builds up, starts with numerical effects. 
This includes, but is not limited to, residual divergence cleaning error, numerical damping, and numerical diffusion. Our simulations employ a combination of the Powell \citep{powell1999solution} and Dedner \citep{dedner2002hyperbolic} divergence cleaning scheme, which controls the relative divergence error to $\lsim 10\%$ during most of the amplification stage (see the right panel of \fig{par_tracing_tracks} and discussion in \se{4.1.1}). However, this only means that the error is dynamically unimportant and does not guarantee the same for the growth of the B-field. Residual errors would be added to the right hand side of \equ{logB evolution}, and if we compute $\left(\frac{d\ln B}{dt}\right)_{\rm kin}$ from $\vec{v}$ and $\vec{b}$ and compare it to the actual $\frac{d\ln B}{dt}$ from the simulations, we could potentially conclude on how much the numerical effect contributes to the rapid growth of B-field (see Appendix. \ref{sec:div cleaning} for detail). More effective divergence cleaning schemes, such as the method with a more globally controlled scheme, the constrained gradient (CG; see \citet{hopkins2016constrained}) could potentially alleviate such effects, and thus affect the results of the field strengths and the evolutionary course. Reruns are needed to understand these impacts, and we plan on running simulations with such variations.

Based on the above discussion, we conjecture the following picture in our simulations: numerical effects from divergence cleaning errors or numerical diffusion initiated/triggered the amplification of the B-field at an early time, which then got picked up by further-developed physical mechanisms such as turbulence and feedback, resulting in a surge of $B$ on the scale of our simulation cells. Even though numerical effects play a role here, this amplification stage still ends with a saturated magnetic-to-turbulent ratio consistent with idealized simulations and produces B-fields consistent with low-$z$ observations, and the onset of amplifications is also consistent with the rapid accretion and feedback, so the physical effects should still be robust. The actual physical picture, which happens below our resolution scale, is unknown to us based on our simulations, and could potentially be explained if further developing sub-grid models or running idealized simulations with higher resolution.

\section{Conclusions and Outlooks}
\label{sec:conclusions}

\subsection{Conclusions}
In this paper, we explore the properties and evolution of B-field in simulated FIRE-2 galaxies. 
Our main conclusions are as follows:

\begin{enumerate}
    \item \textbf{B-field properties at $z=0$:} The B-field in our simulated galaxies are in general turbulent at $z=0$, reaching $\sim \mu\Gauss$ at the galactic scale, though dropping drastically in the CGM. The differences in the B-field between the MHD and CR-CD runs are insignificant for the same halo, and we see little halo-to-halo variations among systems of comparable mass. However, mass-dependence is much more pronounced, and more massive haloes possess stronger B-fields, mostly due to stronger turbulence and higher densities. An overall $B\sim\rho^{1/2}$ trend is observed in all of our galaxies at $z=0$, corresponding to flux-freezing effects in a 2D planar/filamentary geometry.

    \item \textbf{B-field evolution:} By tracing gas particles in our simulations, we see a rough three-stage evolution of the B-field:
    \begin{itemize}
        \item 1. Initial expansion stage ($z\gsim 4$), when both the B-field and density are being diluted by the expansion of the Universe, roughly following $B\sim \rho^{2/3}$. This is when a galaxy has not been formed. A sudden hump in the turbulent components of both magnetic and kinetic energies is observed around $z\sim 7-10$, close to the reionization era, but later goes away. We suspect that this is due to sudden turbulence generated by the reionization. At this stage, B-field is weak and the magnetic-to-turbulent ratio is $\lsim 10^{-8}$;
        \item 2. Rapid amplification stage ($1\lsim z\lsim 4$), when the B-field is being amplified violently with small density changes. This is when structure formation starts to kick in, and major mergers, accretion, and star formation drive feedback. The magnetic-to-turbulent ratio increases, reaching $\sim 1\%$ when close to $z\sim 1$. We conclude the possible origin of such amplification in (iii);
        \item 3. Final saturation stage ($z\lsim 1$), when the B-field grows with the density, roughly following $B\sim \rho^{1/2}$. At this stage, the gas flow around the galaxy is much less violent. Multiple mechanisms could be driving the B-field evolution the same time, including flux-freezing and saturation due to dynamical or energetic equilibria. The magnetic-to-turbulent ratio saturates at $\sim 1\%$ for diffuse gas with $\nH\lsim 1\cmc$, and $\sim 10\%$ for denser and closer-to-disk gas with $\nH\gsim 1 \cmc$.
    \end{itemize}

Trends we observed are broadly robust with halo mass (MW-mass galaxies vs. dwarfs), different physics (CR vs. MHD), resolutions, and feedback schemes (FIRE-2 vs. FIRE-3). However, the aforementioned timings only apply for gas ending up in the galaxy (i.e. never having formed stars). For star progenitors, we still observed similar three-stage evolutionary courses, but the time for each stage get shifted earlier, depending on the star formation times.

    \item \textbf{Origin of the amplification:} To explain the rapid amplification stage we see in the B-field evolution (stage 2, also known to be associated with ``dynamos''), we approach from both physical and numerical perspectives. Known physical processes responsible for such growth could only happen below the scale of our simulation's resolution. We conjecture a turbulence-driven amplification process that are likely triggered by numerical effects. Residual numerical errors from either divergence cleaning or numerical diffusion exist in our simulations, and could be driving the initial instabilities to amplify the B-field. Once established, a rapid surge of the B-field is driven by a violent flow in the gas, including turbulence created by large-scale and small-scale motion, advecting the numerical errors and magnifying such effects. 
\end{enumerate}

\subsection{Outlooks}
Although it is clear that the amplification stage of the B-field is rapid and distinctive, we still have little grasp of its precise physical driving factors. The B-field evolution is a complicated process involving interactions of different parts of the galaxy, and in future works we plan to explore different numerical MHD schemes and cleaning methods, including how different seed field values and divergence cleaning schemes could impact the B-field saturated strengths and evolutionary course.

\section*{Acknowledgements}
YSL thanks Ruediger Pakmor, Kung-Yi Su, and Patrick Diamond for useful discussion and technical help. YSL and DK were supported by the National
Science Foundation (NSF) grant AST-2108314. YSL was also supported by the University of California San Diego Academic Senate Research Grant. PFH acknowledges support from a Simons Investigator Grant. SBP acknowledges support from the Knut and Alice Wallenberg Foundation Fellowship program under grant number 2023.0080.

\section*{Data Availability Statement}
The data supporting the plots within this article are available on reasonable request to the corresponding author. A public version of the GIZMO code is available at \hyperlink{http://www.tapir.caltech.edu/~ phopkins/Site/GIZMO.html}{http://www.tapir.caltech.edu/$\sim$ phopkins/Site/GIZMO.html}. Additional data including simulation snapshots, initial conditions, and derived data products are available at \hyperlink{http://fire.northwestern.edu}{http://fire.northwestern.edu}.


\bibliographystyle{mnras}
\bibliography{references.bib}



\appendix

\section{Miscellaneous Figures}\label{Misc plots}

In this appendix, we present some miscellaneous figures that supplement our main results.

\begin{figure}
    \centering
    \includegraphics[width=0.98\linewidth]{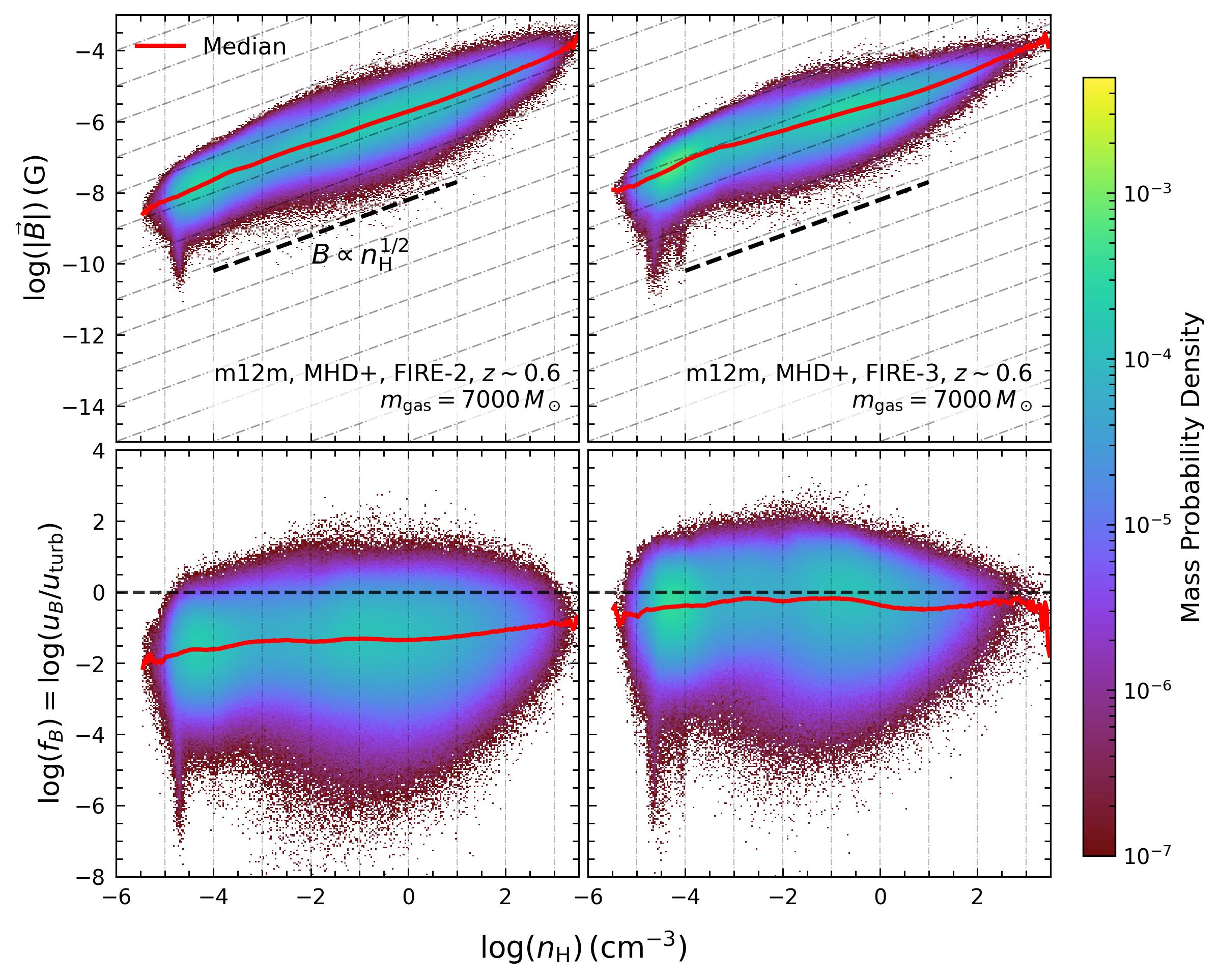}
    \caption{Mass-weighted $B-\nH$ and $f_{\rm B}-\nH$ (see the main text for definition of $f_{\rm B}$) distribution histograms for gas with $r\leq 150\kpc$. We present two runs for another MW-mass galaxy $\texttt{m12m}$: the FIRE-2 high-res MHD run, and the FIRE-3 high-res MHD run. Similar to what we saw in \se{robustness}, going from FIRE-2 to FIRE-3, the B-field strength increases slightly, while the magnetic-to-turbulent ratio $f_{\rm B}$ bumps up much more.}
    \label{fig:robustness_m12m}
\end{figure}

In \fig{robustness_m12m} we present the $B-\nH$ and $f_{\rm B}-\nH$ distribution histograms for gas with $r \leq 150 \kpc$ in two \texttt{m12m} runs, akin to \fig{robustness_B_fB_nH}. This is to support our conclusions in \se{robustness}, in which we only discussed the robustness of basic B-field properties against resolution and feedback schemes for the \texttt{m12i} halo. For the \texttt{m12m} haloes, the FIRE-3 feedback schemes still result in a slightly higher B-field value for a given density, but the magnetic-to-turbulent ratio $f_{\rm B}$ increases more significantly. This is in agreement with what we found in \se{robustness}. We therefore conclude that feedback schemes have a similar impact on all MW-mass galaxies. 

\begin{figure*}
    \centering
    \includegraphics[trim={0.0cm 0.0cm 0.0cm 0.0cm}, clip, width =0.99\textwidth]{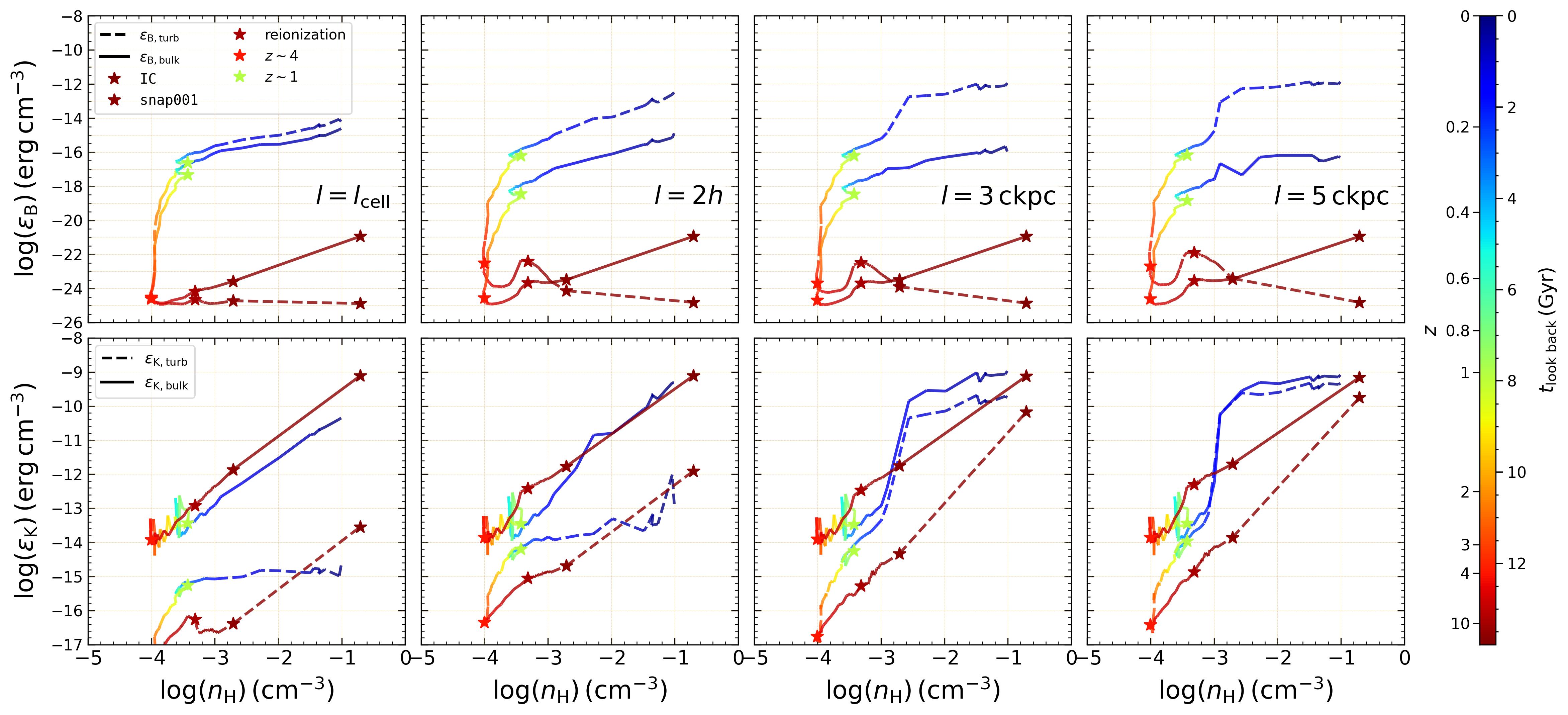}
    \caption{Robustness tests on different filter sizes applied in our decomposition exercise of the magnetic and kinetic energy densities. Each column represents one filter size, with the top panel presenting the magnetic energies and the bottom panel presenting the kinetic energies. Our fiducial result in the main text adopts $l=h$ (the smoothing length), and here we present four more trials: $l=l_{\rm cell}$ (cell size), $l=2h$, $l=3 {\rm \, ckpc}$, and $l= 5 {\rm \, ckpc}$. We see that our results in general converge.}
    \label{fig:filter_sizes_all}
\end{figure*}

In \fig{filter_sizes_all} we present our results of using different filter sizes to decompose the magnetic and kinetic energy densities, verifying the robustness of such choices. These should be compared to our fiducial trial in \fig{bulk_and_turb_fields}. Different filter sizes were labelled at the right side of each top panel. We see that with different filter sizes, although the values of different components may change (particular at low-$z$), the general evolutionary trends are robust and independent of the filter sizes. When using the two constant filters ($l=3$ and $5$ ckpc), we ``smear'' over large regions of relatively high-density galactic gas, causing a more significant change in $\varepsilon_{\rm K}$. Interestingly, with all different applied filters, we see the same humps in both the magnetic and kinetic energies around the reionization era. The impact of different filter sizes is stronger on the kinetic field than on the magnetic field. This is because turbulence happens across all scales so shifting the filter size would cause changes in the turbulence strength. A more detailed discussion on different turbulent decompositions and their implications would be beyond the scope of this study.

\begin{figure*}
    \centering
    \includegraphics[trim={0.0cm 0.0cm 0.0cm 0.0cm}, clip, width =0.95\textwidth]{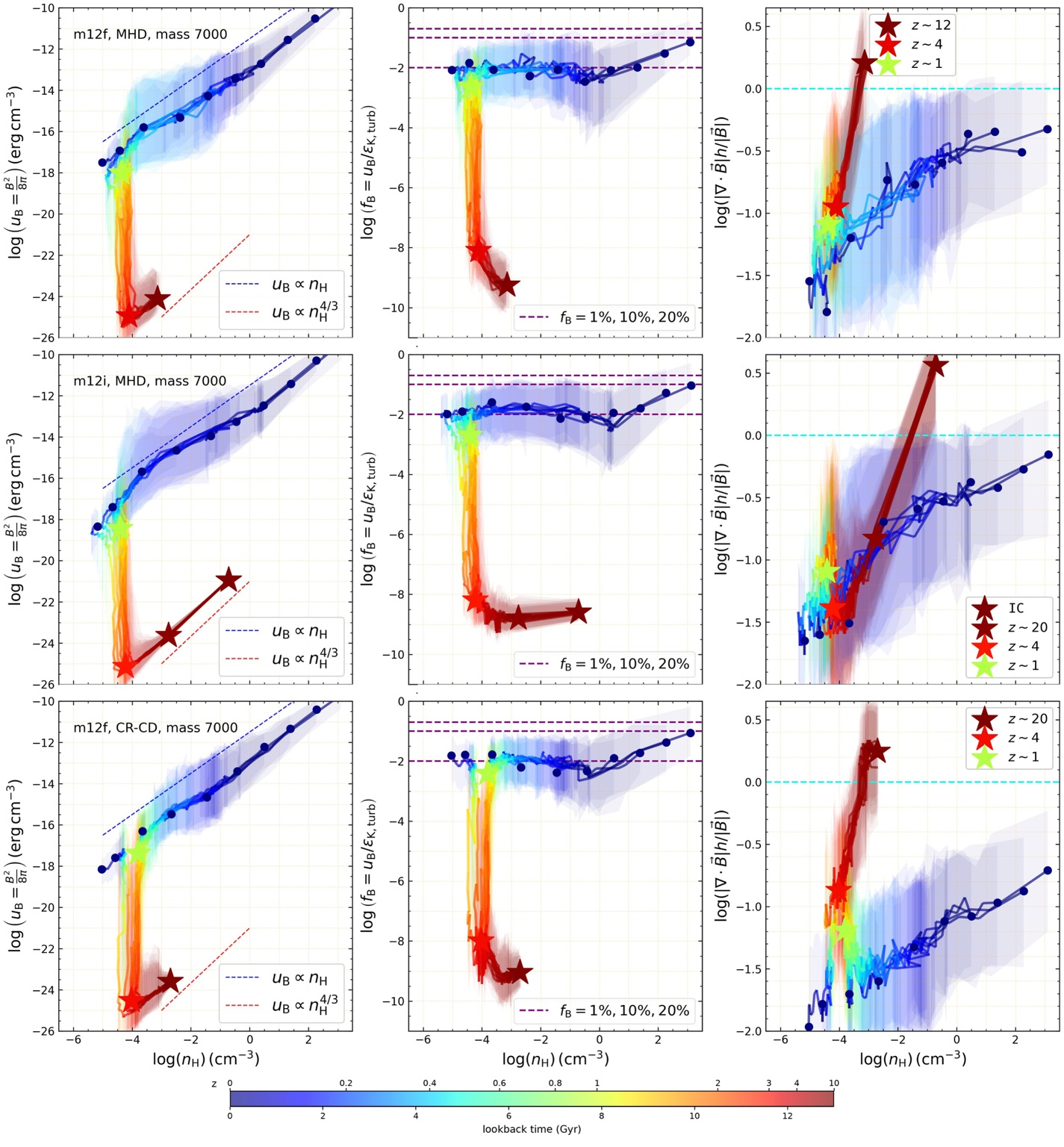}
    \caption{Particle tracing tracks for the rest of our MW-mass galaxy runs. See \fig{par_tracing_tracks} for detailed description of the panels. Each row shows the result for one simulation run. The important milestone times discussed in the main text are also labelled, although due to availability of the snapshots some runs start from either $z\sim 20$ or $12$, so the starting points vary. For the same reason, the tracks for the \texttt{m11b} runs appear to be less smooth. Nevertheless, similar features are observed, and the magnetic-to-turbulent ratios all reach $\sim 1-10\%$ regardless of the size of the galaxy or the physics included (CR vs. MHD).}
    \label{fig:misc_tracks_MWs}
\end{figure*}

\begin{figure*}
    \centering
    \includegraphics[trim={0.0cm 0.0cm 0.0cm 0.0cm}, clip, width =0.92\textwidth]{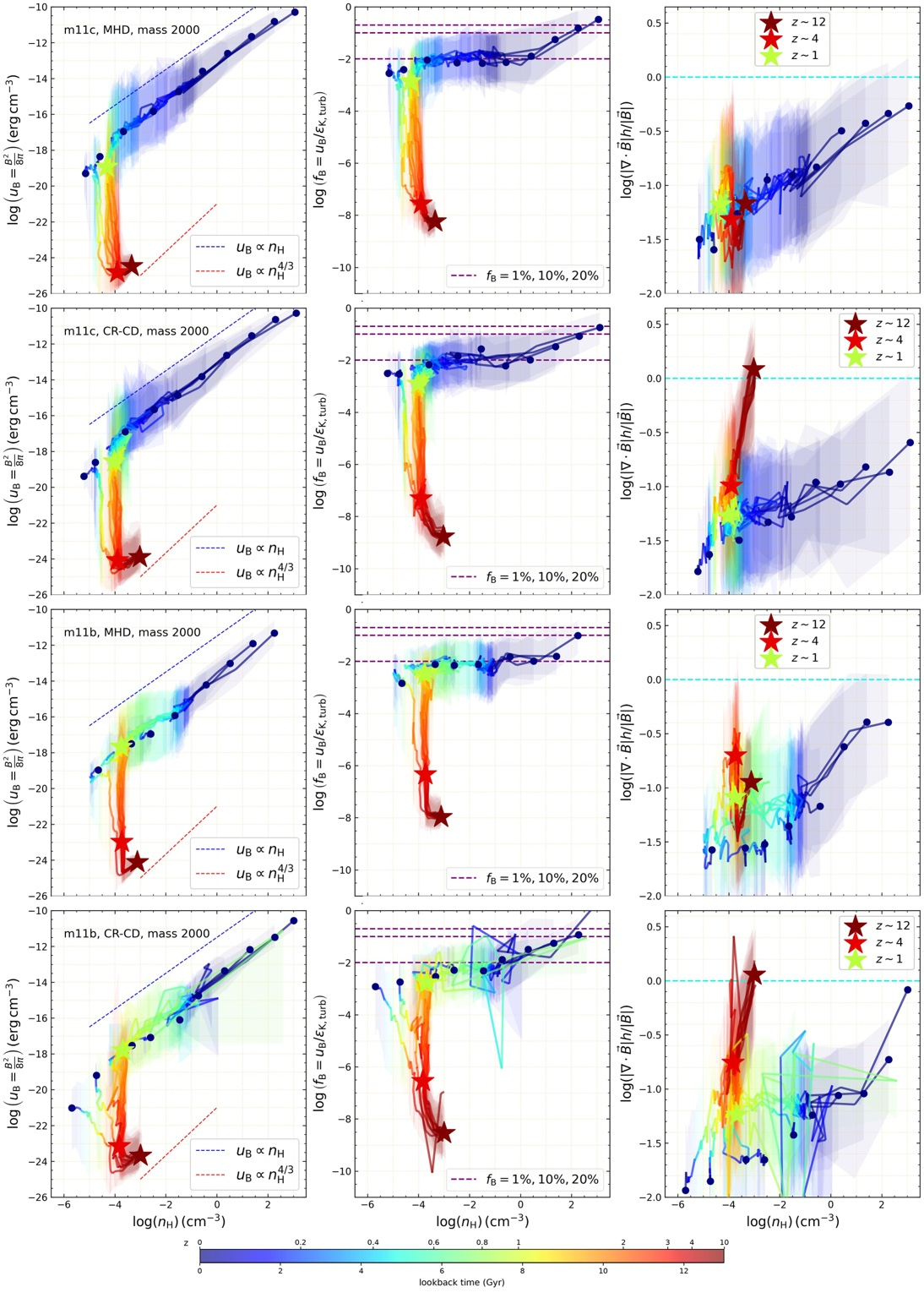}
    \caption{Same as \fig{misc_tracks_MWs}, but for our dwarf runs. We again see robust features among all presented runs.}
    \label{fig:misc_tracks_dwarfs}
\end{figure*}

In \figs{misc_tracks_MWs} and \figss{misc_tracks_dwarfs} we present the particle tracing tracks for different properties for all of our other galaxies. All arrangements and presentation are the same as for our fiducial run ($\texttt{m12i}$ CR-CD) in the main text (\fig{par_tracing_tracks}). We see that conclusions drawn from our fiducial run apply for all other runs, making them robust features in galactic B-field evolutions. A three-stage evolution is seen in all of our runs. Importantly, the magnetic-to-turbulent ratios saturate around similar values ($\sim 1-10\%)$ for $z\lsim 1$ even with different halo sizes, which confirms the role of turbulence played in amplification, rather than being amplified randomly by numerical effects. The ``dip-then-rise'' features for high-density gas are also seen in all our runs, even for the $\texttt{m11c}$ halo, which is an irregular galaxy without a compact disk. With a less massive halo come weaker B-fields, but also with weaker turbulence. The fact that the magnetic-to-turbulent ratios are maintained constant indicates the passivity of B-field, as the saturated strengths are controlled by the turbulence strengths. 

\section{Divergence cleaning}\label{sec:div cleaning}


In \se{discussion} we conjecture a possibility that the amplification may be induced by numerical effects, and one of the major numerical effects is the residual error from divergence cleaning. In this appendix, to aid our conjecture, we provide more insights by schematically showing the actual evolution of B-field in our simulations. 

The actual induction equation that is evolved in our simulations with additional source terms aiming to ``clean'' the non-zero divergence \citep{hopkins2016accurate}:

\be\label{eq:actual induction}
    \pdv{\vec{B}}{t}-\nabla\times(\vec{v}\times\vec{B})=-\underbrace{\vec{v}(\nabla\cdot\vec{B})}_{\rm Powell}-\underbrace{\nabla\psi}_{\rm Dedner}.
\ee
Here, the first term corresponds to the cleaning employed by the Powell method \citep{powell1999solution}, which advects back the non-zero divergence along the magnetic field lines. The second term corresponds to the Dedner method \citep{dedner2002hyperbolic}, which is an additional damping term to smooth out the consequences from the Powell method. Note that the Powell term is simply explicitly subtracting the term that would have been $\propto \nabla\cdot\vec{B}$ when expanding the triple product in the induction equation. With the additional source terms and dropping the previous assumption on $\nabla\cdot\vec{B}=0$, \equ{logB evolution} is then modified to

\begin{align}\label{eq:dlnBdt continuum}
    \frac{d\ln B}{dt}=\underbrace{\vec{b}^T(\nabla\vec{v})\vec{b}-\nabla\cdot\vec{v}}_{\equiv \left(\frac{d\ln B}{dt}\right)_{\rm kin}}
    &+ \underbrace{\frac{\vec{v}\cdot\vec{b}}{B}\nabla\cdot\vec{B}}_{\rm the \, divergence \, error} \notag \\
    & -\underbrace{\frac{\vec{v}\cdot\vec{b}}{B}\nabla\cdot\vec{B}}_{\rm Powell}
    - \underbrace{\frac{\vec{b}}{B}\cdot\nabla\psi}_{\rm Dedner}.
\end{align}

\no In the continuum limit, the divergence-error term and the Powell term in
\equ{dlnBdt continuum} cancel algebraically. However, this is not necessarily true numerically. To see this, for a fluid element
$i$, the actual MFM update can be schematically written as
\be
    \left(\frac{\Delta \ln B}{\Delta t}\right)_i
    =
    \left(\frac{\Delta \ln B}{\Delta t}\right)_{{\rm kin},i}
    +
    \Gamma_{{\rm clean},i},
\ee

\no where the kinetic contribution is

\be
    \left(\frac{\Delta \ln B}{\Delta t}\right)_{{\rm kin},i}
    \equiv
    \vec{b}_i^T
    \left(\nabla\vec{v}\right)_i
    \vec{b}_i
    -
    \left(\nabla\cdot\vec{v}\right)_i ,
\ee

\no and

\begin{align}
    \Gamma_{{\rm clean},i}
    \equiv
    \frac{\vec{b}_i}{B_i}\cdot
    \left[
        \vec{v}_i
        \left(\nabla\cdot\vec{B}\right)_{{\rm ind},i}
        -
        \vec{v}_i
        \left(\nabla\cdot\vec{B}\right)_{{\rm P},i}
        -
        \left(\nabla\psi\right)_{{\rm D},i}
    \right]
    + \mathcal{O}(\Delta t,h_i^p)
\end{align}

\no is the residual divergence cleaning error. Here, $\left(\nabla\cdot\vec{B}\right)_{{\rm ind},i}$ denotes the effective
discrete monopole contribution entering the induction equation, while
$\left(\nabla\cdot\vec{B}\right)_{{\rm P},i}$ denotes the corresponding
discrete quantity entering the Powell source term. These two quantities represent the same continuum $\nabla\cdot\vec{B}$, but in the numerical scheme they are evaluated through reconstructed states and meshless gradient/divergence estimators. Therefore, they may not be identical at finite resolution, even though they become identical in the continuum limit. The residual term $\Gamma_{{\rm clean},i}$ thus represents
the finite-resolution cleaning error that may contribute to the instantaneous rapid growth of $B$ seen in the main text results.

\bsp	
\label{lastpage}
\end{document}